\documentclass[a4paper,fleqn]{cas-dc}

\usepackage[numbers,sort&compress]{natbib}
\usepackage[utf8]{inputenc}
\graphicspath{{../pdf/}{../jpeg/}}
\DeclareGraphicsExtensions{.pdf,.jpeg,.png}
\usepackage{array}
\usepackage{url}
\usepackage[center]{caption}
\usepackage{algorithmicx}
\usepackage{algpseudocode}
\usepackage{mathtools}
\usepackage{booktabs}
\usepackage{rotating}
\usepackage{lscape}
\usepackage{comment}
\usepackage{caption}
\usepackage{multirow}
\usepackage[ruled,vlined]{algorithm2e}
\usepackage{empheq} 
\usepackage{appendix}
\usepackage{threeparttable}  
\usepackage{graphicx}
\usepackage{float}
\usepackage{amsmath}
\usepackage{epstopdf}
\AppendGraphicsExtensions{.tif}
\usepackage{subfigure}

\usepackage{booktabs,caption}
\usepackage{diagbox, eqparbox, hhline}

\usepackage{tabulary}
\usepackage{placeins}

\def\tsc#1{\csdef{#1}{\textsc{\lowercase{#1}}\xspace}}
\tsc{WGM}
\tsc{QE}

\begin{document}
\let\WriteBookmarks\relax
\def\floatpagepagefraction{1}
\def\textpagefraction{.001}

\shorttitle{HybridGuard: Enhancing Minority-Class IDS}    

\shortauthors{Kar et al.}

\title [mode = title]{HybridGuard: Enhancing Minority-Class Intrusion Detection in Dew-Enabled Edge-of-Things Networks}

%







\author[1]{Binayak Kar} 
\cormark[1]
\ead{bkar@mail.ntust.edu.tw}

\affiliation[1]{organization={Department of Computer Science and Information Engineering, National Taiwan University of Science and Technology}, city={Taipei}, country={Taiwan}}

\author[1]{Ujjwal Sahu}

\author[2]{Ciza Thomas}
\affiliation[2]{organization={Department of Computer Science and Engineering, Digital University Kerala}, city={Thiruvananthapuram}, state={Kerala}, country={India}}

\author[1]{Jyoti Prakash Sahoo}




\cortext[1]{Corresponding author}

\begin{abstract}
Securing networks in Dew-Enabled Edge-of-Things (EoT) networks from sophisticated intrusions is a challenge that is at once critical and challenging. This paper presents HybridGuard, a state-of-the-art framework that combines Machine Learning and Deep Learning to raise the bar for intrusion detection. HybridGuard addresses data imbalance by performing mutual information-based feature selection to ensure that the most important features are always considered to improve detection performance, especially for minority attacks. The proposed framework leverages Wasserstein Conditional Generative Adversarial Networks (WCGAN-GP) to alleviate class imbalance, hence enhancing the precision of detection.
In the framework, a two-phase architecture named  
``DualNetShield" was integrated to introduce advanced network traffic analysis and anomaly detection techniques, enhancing the granular identification of threats within complex EoT environments. HybridGuard, tested on UNSW-NB15, CIC-IDS-2017, and IOTID20 datasets, demonstrates robust performance over a wide variety of attack scenarios, outperforming the existing solutions in adaptation to evolving cybersecurity threats. This innovative approach establishes HybridGuard as a powerful tool for safeguarding EoT networks against modern intrusions.
\end{abstract}


\maketitle

\section{Introduction}
\label{sec:Intro}
{T}{he} rapid expansion of internet usage, alongside the proliferation of Internet of Things (IoT) devices and advancements in cloud and edge computing, has significantly heightened the risk of cyber-attacks across diverse networks. IoT devices, particularly those used in home automation, environmental monitoring, and smart mobility, are especially vulnerable due to their low security standards and the reluctance of manufacturers to provide timely software updates. This negligence exposes outdated software to exploitable vulnerabilities, posing severe risks to financial and reputational security. Moreover, with the increasing usage of IoT devices and the intuitive design of IoT networks, various security threats and challenges have emerged, creating behavioral uncertainty in the Internet \cite{thakkar2021attack}.

Additionally, as IoT solutions and high-performance applications become more prevalent, the cloud model alone faces challenges in meeting new demands. Although cloud computing \cite{yu2017survey} offers strong, networked resources like storage and remote processing, it struggles to efficiently support the migration of virtual resources and sustain the productivity of IoT devices during data transfers. 
As highlighted in \cite{el2017edge}, edge computing has emerged as a complementary framework to cloud systems, referred to as Edge-of-Things (EoT) computing, to address these limitations.
EoT enables the processing and storage of information closer to IoT devices, allowing near real-time data handling, which is essential for applications such as live video streaming, gaming, and industrial monitoring. By seamlessly transferring services and data processing from the cloud to edge devices, EoT enhances responsiveness in large-scale IoT ecosystems, such as industrial networks. However, this shift also introduces new privacy and security challenges, as the distributed EoT environment requires robust intrusion detection systems to protect these networks from potential threats.

Network Intrusion Detection Systems (NIDS) are pivotal in defending against these evolving cyber threats, continuously monitoring network activity to detect and respond to intrusions \cite{sahoo2025qcshield, he2023adversarial}. However, their effectiveness is often hampered by the class imbalance problem inherent in network log data. This imbalance, where attack instances are vastly outnumbered by normal traffic, skews the model’s performance, leading to poor detection of minority attacks, which are often the most critical and sophisticated threats \cite{thakkar2020role}.

The datasets UNSW-NB15 \cite{moustafa2015unsw}, CIC-IDS2017 \cite{sharafaldin2018toward}, and IOTID20 \cite{ullah2020scheme} exemplify this challenge, where the imbalance between normal and attack classes results in high false positive rates and reduced detection accuracy for minority attacks. Traditional approaches, such as the Synthetic Minority Oversampling Technique (SMOTE) \cite{chawla2002smote, elreedy2019comprehensive} and cost-sensitive algorithms, attempt to address this imbalance but often fall short in achieving the required precision, recall, and overall performance balance. Effective feature selection is also essential to improving NIDS performance by reducing overfitting and enabling models to focus on the most relevant features \cite{ahmed2016survey}. Techniques such as genetic algorithms and particle swarm optimization have been explored in this regard, yet they remain insufficient to address the full scope of challenges posed by modern network environments.


With the growing influence of deep learning (DL) techniques, extensive research has focused on developing intrusion detection systems (IDS) capable of analyzing and classifying network traffic as either normal or malicious \cite{sahoo2026choir, ferrag2020deep}. Established DL approaches have shown intricate learning capabilities and adaptability across various application domains \cite{goodfellow2016deep}, \cite{dong2016comparison}. However, the effectiveness of these techniques is often hindered when training data samples are insufficient or highly disproportionate among class output labels \cite{duy2023investigating}. Such imbalances lead to the generation of a high number of false positives and low detection capability, as the models tend to focus disproportionately on the majority class, neglecting minority classes \cite{liu2020privacy}, \cite{thakkar2021review}.

To bridge these critical gaps, we present ``HybridGuard,'' an advanced framework that integrates Machine Learning and Deep Learning to strengthen intrusion detection capabilities.
Unlike traditional methods, HybridGuard employs Wasserstein Conditional Generative Adversarial Networks (WCGAN-GP) to effectively mitigate data imbalance, enhancing the model's ability to detect minority attacks without inflating false positives. 
Furthermore, its mutual information-based feature selection method surpasses traditional approaches by systematically identifying the most relevant features, ensuring improved detection granularity while minimizing overfitting.
HybridGuard’s unique integration of DualNetShield, ``a two-phase architecture to enhance intrusion detection by addressing both major and minor attack classes,'' provides advanced network traffic analysis and anomaly detection methods, optimizing the detection of sophisticated and evolving attack vectors. 
Experimental evaluations on UNSW-NB15 \cite{moustafa2015unsw}, CIC-IDS2017 \cite{sharafaldin2018toward}, and IOTID20 \cite{ullah2020scheme} datasets demonstrate HybridGuard's superior performance across a range of attack scenarios.

\textbf{Research Gaps:} 
There are three primary gaps in existing research on NIDS: 
\begin{enumerate}
    \item Most studies primarily target the majority attack classes, which have a larger number of samples or occur more frequently, neglecting the less frequent yet critical minority attacks.
    \item The issue of class imbalance in datasets is often overlooked, leading to poor detection rates for minority attack classes.
    \item Many approaches suffer from a high False Alarm Rate (FAR), resulting in unnecessary computational load on network administrators.
\end{enumerate}

\textbf{Contributions:} 
To address these research gaps, this paper introduces the HybridGuard framework, which significantly improves NIDS capabilities. The key contributions of this work are as follows:

\begin{enumerate}
    \item Proposes a novel HybridGuard framework to enhance Network Intrusion Detection Systems, specifically focusing on identifying and categorizing minority attacks.
    \item Implements a class imbalance solution using WCGAN-GP, generating synthetic data to balance datasets 
    and improving the overall detection rate.
    \item Enhances minority attack identification by employing a mutual information-based feature selection method, which better detects stealthy, less frequent attacks.
    \item Integrates DualNetShield within the framework, leveraging advanced network traffic analysis and anomaly detection with a specific focus on identifying minority attacks.
    \item By improving minority attack detection, the HybridGuard framework ensures better protection against subtle yet potentially devastating intrusions, reducing the risk of data breaches and enhancing overall network security. 
\end{enumerate}

The rest of the paper is organized as follows. The related works are discussed in Section~\ref{sec:related_work}. 
We discuss the background information and our motivation towards this research focus in Section~\ref{sec:background_motivation}.
We introduce our proposed HybridGuard framework in Section~\ref{sec:proposed_methodology}. 
In Section~\ref{sec:experiment}, we explain the evaluation setup and dataset, along with the data description and hyperparameter tuning model training steps. The results analysis is done in Section~\ref{sec:result}. 
Finally, we conclude the paper in Section~\ref{sec:conclusion}.


\section{Related Works}
\label{sec:related_work}

Recent research in Network Intrusion Detection Systems (NIDS) focuses on solving the issue of data imbalance to improve the detection of minority attacks. Models like CWGAN-CSSAE use Conditional Wasserstein GANs to generate synthetic samples of minority attacks and cost-sensitive learning to enhance detection accuracy \cite{zhang2020network}. These methods, along with others like GAN-RF and SAVAER-DNN \cite{yang2020network}, have shown significant improvements in identifying rare attacks by addressing class imbalance and applying deep feature extraction techniques.

Identifying unknown threats and managing class imbalance are two critical challenges in intrusion detection for IoT networks. Telikani et al. \cite{telikani2023cost} addressed these issues by integrating multitask learning with cost-sensitive learning. They employed support vector machines and stacking autoencoders to improve the detection of novel and low-frequency intrusions. Zeng et al. \cite{zeng2024causal} introduced the Causal Genetic Network Anomaly Detection (CNSGA) framework, which combines genetic algorithms with causal inference to enhance detection rates.

Singh et al. \cite{singh2020daas} introduced a DaaS (Dew Computing as a Service) framework to enhance intrusion detection in Edge-of-Things (EoT) networks. The framework combines deep learning, particularly a Deep Belief Network (DBN), with dew computing, which effectively balances local and cloud resources for more efficient processing. This integration reduces communication delays and cloud server workload, addressing the challenges of traditional intrusion detection systems at the edge. 
To improve IDS performance, Ullah et al. \cite{ullah2022new} prioritized deep learning (DL) strategies like CNNs and stressed the drawbacks of conventional ML techniques. A Self-Attention-Based Deep CNN (SA-DCNN) that lowers computational overhead and boosts accuracy was proposed by Alshehri et al. \cite{alshehri2024self}. Furthermore, ADCL \cite{ma2023adcl}, a collaborative learning framework, achieves notable F-score gains by utilizing several models to improve detection adaptability, integrity, and capacity. These methods demonstrate how IDS has advanced for IoT security. 

The growing intricacy of networks, especially with developments like 5G, makes it difficult for conventional IDS to handle class imbalance, which reduces the accuracy of minority attack detection. Su et al. \cite{su2020bat} used bidirectional LSTM to improve detection in imbalanced datasets. LIO-IDS was proposed by Gupta et al. \cite{gupta2021lio}, who used LSTM with I-OVO and Borderline-SMOTE to improve multi-class detection and handle imbalanced datasets. Ding et al. \cite{ding2022imbalanced} improved detection metrics like ``accuracy'' and ``recall'' on real-world datasets by combining TACGAN oversampling for attacks with KNN undersampling for normal data.
Wisanwanichthan et al. \cite{wisanwanichthan2021double} presented a Double-Layered Hybrid Approach (DLHA) for Network Intrusion Detection Systems (NIDS), combining Naive Bayes and Support Vector Machine (SVM) classifiers.
The DLHA model addresses the challenge of detecting uncommon attacks more effectively than single machine learning models, showing that combining classifiers can significantly improve detection accuracy.

The use of GANs in CPS domains for attack scenarios and robust system design has been investigated in recent publications. Metrics like Jensen–Shannon or Kullback–Leibler divergence \cite{kim2017malware,shahriar2020g,yilmaz2020addressing} were used in early GAN-based research \cite{goodfellow2014generative}, whereas more sophisticated versions like Wasserstein GANs \cite{li2020improving,lee2019generation} and conditional GANs \cite{dlamini2021dgm} aimed at raising the performance of minority classes. Li et al. \cite{li2018anomaly} presented an LSTM-GAN in industrial settings, whereas Alabugin et al. \cite{alabugin2020applying} created a bidirectional GAN for anomaly detection that was verified on the SWaT dataset. Siniosoglou et al. \cite{siniosoglou2021unified} showed efficacy in smart grid situations by integrating an autoencoder within GANs for anomaly detection and attack categorization. 
Unlike previous GAN approaches based on data distribution distances, our work emphasizes a reconstruction error-based GAN model to generate more realistic synthetic data. Specifically, we utilized the concepts with the Wasserstein distance between the reconstruction error distributions of real and synthetic samples. 


\section{Background and Motivation} \label{sec:background_motivation}
The increasing complexity of modern networks, coupled with the rise of sophisticated cyber threats, underscores the critical need for robust NIDS. 
However, traditional NIDS often struggle with detecting minority attacks, which, despite their rarity, can have severe consequences. 
This section discusses the security challenges of NIDS, the significance of focusing on minority attack detection, and provides an overview of the GAN-based techniques employed in our solution. 
Finally, we present the motivation for this research, highlighting the need to advance NIDS to address underrepresented but highly damaging threats more effectively.

\subsection{Security Challenges in NIDS}

NIDS faces several critical challenges, including high false positive rates that overwhelm administrators with unnecessary alerts, and class imbalance in datasets, which reduces accuracy for underrepresented attack types \cite{abdulganiyu2023systematic}. Implementing advanced NIDS can also be resource-intensive, limiting their feasibility in real-time or constrained environments. Additionally, evolving cyber-attack patterns, especially zero-day attacks and minority attacks, make it difficult for NIDS to keep up, while the rise of encrypted traffic complicates threat detection. Moreover, selecting relevant features for detection is challenging, as irrelevant data can lower accuracy and increase processing time. These issues underscore the need for more adaptable, efficient NIDS that can meet the demands of modern cybersecurity.

\begin{figure}[!t]
\centering
\includegraphics[width=0.5\textwidth, height=0.39\textheight]{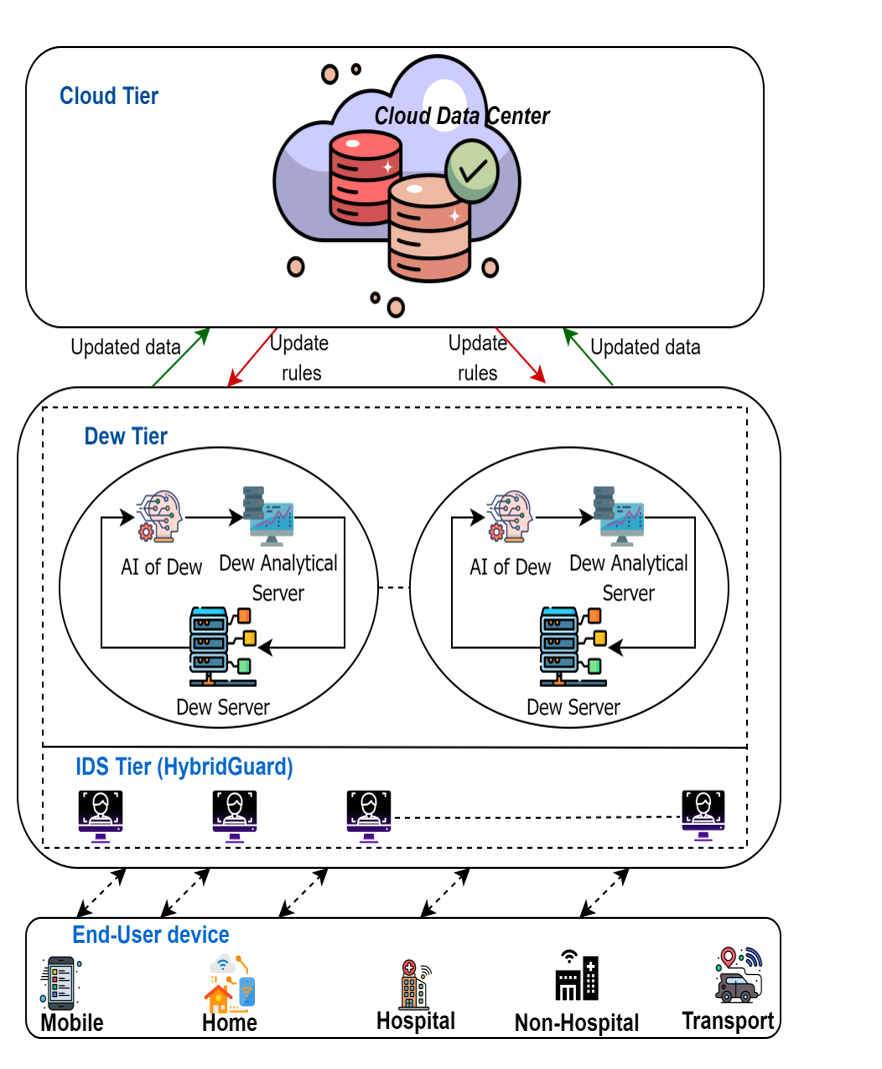} 
\vspace{-12pt} 
\caption{Dew-enabled Edge-of-Things networks.} 
\label{fig:Dew} 
\vspace{-4pt} 
\end{figure}


\subsection{Dew Computing for Intelligent IDS}

Dew computing is centered on two key ideas: Collaboration \cite{wang2016definition}, and Independence \cite{ristov2016implementation}. It enables local devices like smartphones and laptops to process data near its source, reducing the need for centralized cloud systems. This approach aims to make network management easier and more efficient by using on-site computing resources.
Dew computing is characterized by six main features: Accessibility, Transparency, Re-origination, Scalability, Synchronization, and Rule-based data collection. It supports models like Infrastructure-as-a-Dew (IaaD) and Software-as-a-Dew (SaaD), which help reduce the constant dependence on cloud services. This setup allows for better energy efficiency and less communication delay.

In the realm of Intelligent Intrusion Detection Systems (IIDS), dew computing improves processing times and lessens the load on networks by handling data locally before sending it to the cloud. 
A network enabled by dew is presented in Figure \ref{fig:Dew}, which includes three layers:

\begin{itemize}
    \item \textbf{Cloud-tier:} Handles heavy computations and data processing for IDS. It processes data that has been pre-processed by the dew-tier before uploading it to the cloud servers, reducing communication overhead.
    \item \textbf{Dew-tier:} Equipped with essential software, this tier focuses on selecting algorithms, enabling quick local decision-making with minimal delay. It stores traffic data temporarily and uploads it to the cloud at user-defined intervals.
    \item \textbf{IDS-tier:} This layer performs intrusion detection using data from various nodes. It adapts by selecting algorithms to run either on the dew-tier or cloud-tier, depending on the situation.
\end{itemize}
The system decides whether to use local (Dew) or cloud services based on the current setup, with results being sent to the cloud for more detailed analysis.

\subsection{Generative Adversarial Networks} Generative Adversarial Network (GAN), introduced by Goodfellow et al. \cite{goodfellow2014generative} in 2014, is a significant advancement in deep learning. They consist of a two-player game; these networks are trained concurrently so that each network gains from the actions of the other. While the generator seeks to create synthetic samples that closely resemble the real data, the discriminator seeks to discern between real and synthetic (false) samples.
The ``real'' samples are the actual data from the dataset, while the ``fake'' samples are the artificial data created for the minority classes during the GAN training process. 
Notations used in this paper are listed in Table \ref{tab:parameters_description} with their descriptions.

There are several iterations of the training. 
Every iteration begins with training the discriminator to update its discriminator loss parameters, 
then moves on to training the generator to update its generator loss parameters. 
This process is repeated for a predetermined number of iterations. 
Compared to the generator, the discriminator is often trained more frequently during each iteration. 
Batches of synthetic data (labeled as `0') and real data (labeled as `1') produced by the generator are fed into the discriminator during training. The weights of the discriminator are modified by backpropagating its mistakes.
Jensen-Shannon (JS) divergence is used by the loss function of a simple GAN to calculate the difference between the generated and real data distributions. 



\begin{table}[!t]
\centering
\caption{Parameters used and their description}
\label{tab:parameters_description}
\begin{tabular}{|p{1.35cm}|p{6.1cm}|}
\hline
\textbf{Parameters} & \textbf{Description} \\ \hline
N & Number of training iterations \\ \hline
K & Discriminator extra steps \\ \hline
M & Minibatch size \\ \hline
$\theta_d$ & Parameter of discriminator loss \\ \hline
$\theta_g$ & Parameter of generator loss \\ \hline
$V(D, G)$ & Value function that captures the adversarial relationship between the generator and the discriminator. \\ \hline
$x$ & Original samples from training dataset $\{x_1,x_2,\dots,x_n\}$ \\ \hline
$z$ & Random noise samples $\{z_1,z_2,\dots,z_n\}$ \\ \hline
$G(z)$ & Generated (synthetic) sample \\ \hline
$P_{\text{data}}(x)$ & Probability distribution of the real (actual) data \\ \hline
$P_{\text{data}}(z)$ & Probability distribution of the noise vector \\ \hline
$P_r$ & Distribution of real data \\ \hline
$P_g$ & Distribution of generated data  \\ \hline
$\gamma_(x, y)$ & Represents the joint distribution of \(x\) and \(y\) in \(P_r\) and \(P_g\) \\ \hline

\end{tabular}
\end{table}




Conditional GAN (CGAN) \cite{mirza2014conditional} extends this framework by incorporating additional information \( y \) (such as class labels) into both the generator and discriminator. 
However, CGAN still faces challenges like mode collapse and unstable training.

%



The Wasserstein GANs (WGAN) \cite{arjovsky2017wasserstein} follow the same approach as the standard GAN; the only difference is the loss function. It uses Wasserstein distance 
as the loss function. This distance provides a more consistent statistic by indicating the amount of ``work'' required to convert one distribution into another.
It leads to a more stable training process. It removes the vanishing gradient problem and the mode collapse problem.


In Wasserstein Generative Adversarial Network with Gradient Penalty (WGAN-GP) \cite{gulrajani2017improved}, a gradient penalty is added to the loss function of the WGAN.
By employing the gradient penalty to enforce Lipschitz continuity and the Wasserstein distance as the goal function, Wasserstein GAN with Gradient Penalty (WGAN-GP) increases the training stability of GANs. This leads to more flexibility in the network architecture design, improved mode coverage, and higher-quality samples.

\subsection{Research Motivation}

The motivation for this research stems from the critical need to improve NIDS in detecting minority attacks, which are often overshadowed by more frequent threats. Traditional approaches often focus on majority attack types, leading to poor detection rates for less frequent but equally harmful intrusions. Additionally, the class imbalance in existing datasets further exacerbates this issue, making it difficult to accurately identify minority attacks. High False Alarm Rates (FAR) in current methods also pose challenges by burdening network administrators with unnecessary alerts.

Motivated by these challenges, this research introduces the HybridGuard framework, designed to enhance the detection and classification of minority attacks within NIDS. By incorporating innovative techniques like WCGAN-GP for class imbalance resolution, mutual information-based feature selection, and the DualNetShield for advanced anomaly detection, this work aims to significantly improve overall system accuracy while reducing false alarms.


\section{Proposed HybridGuard Framework} 
\label{sec:proposed_methodology}

%

\begin{figure}[!t]
\centering
\includegraphics[width=0.5\textwidth, height=0.39\textheight]{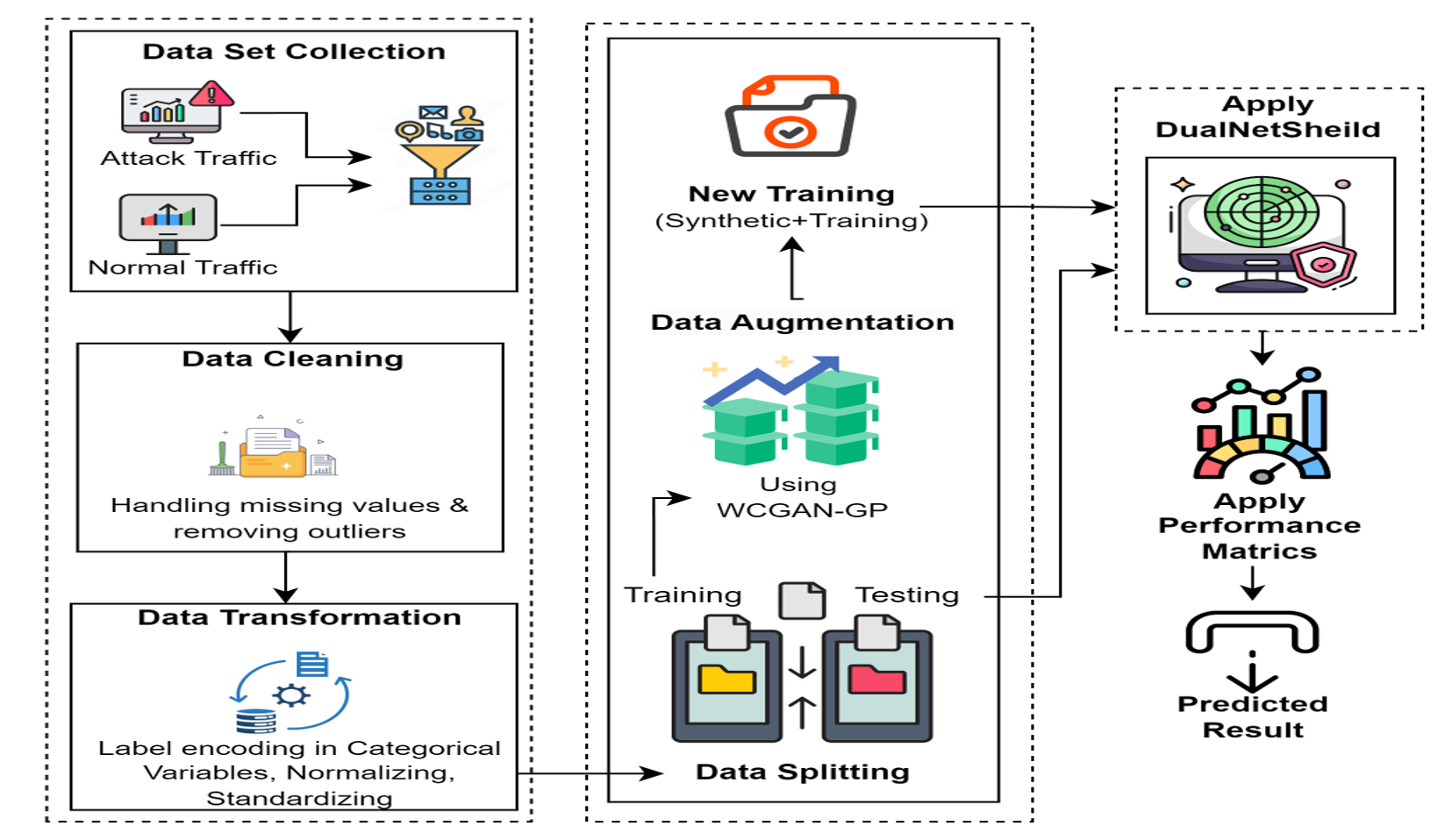} 
\vspace{-12pt} 
\caption{Proposed HybridGuard framework.} 
\label{fig:hybridguard} 
\vspace{-4pt} 
\end{figure}

In this section, we discuss the working principle of our proposed framework in detail, with advancing intrusion detection in Dew-enabled edge-of-things (EoT) environments with the HybridGuard model. The proposed framework has three main modules, as depicted in Figure \ref{fig:hybridguard}: (1) Dataset Selection \& Collection, (2) WCGAN-GP-based Data Oversampling, and (3) DualNetShield. 


\subsection{Data Collection}
In this work, to evaluate the effectiveness of the proposed framework, we conducted experiments using well-known benchmark datasets—UNSW-NB15 \cite{moustafa2015unsw}, CIC-IDS2017 \cite{sharafaldin2018toward}, and IOTID20 \cite{ullah2020scheme}—that are often utilized in intrusion detection research are used in this work. Both typical network traffic and attack types are present in these datasets. 
Each dataset is briefly described in Table \ref{tab:table-2}.

\textbf{UNSW-NB15:} The University of New South Wales's Moustafa et al. \cite{moustafa2015unsw} generated the UNSW-NB15 dataset to solve problems with earlier datasets such as NSL-KDD. Approximately 2.5 million examples were gathered over the course of 31 hours, utilizing a variety of techniques. 43 features total—39 numerical and 4 nominal—are grouped into basic, flow, time, content, and supplementary features in the dataset. The two most important properties are ``attack\_cat,'' which indicates the kind of attack, and ``label,'' which indicates if the traffic is legitimate or malicious.

\textbf{CIC-IDS2017:} The Canadian Institute for Cybersecurity (CIC) provides the CICIDS2017 dataset  Sharafaldin et al. \cite{sharafaldin2018toward}, which is frequently used by researchers to test and create novel intrusion detection models. Compared to the NSL-KDD dataset, which covers a smaller number of protocols and attack methods, CICIDS2017 is a more recent and extensive dataset. It has fifteen distinct forms of traffic, including BENIGN, Heartbleed, and several DDoS attacks, and seventy-eight features.

\textbf{IoTID20:} The IoTID20 dataset \cite{ullah2020scheme} was created by Ullah and Mahmoud to address the need for robust intrusion detection systems tailored to IoT networks. It simulates a smart home network environment consisting of legitimate IoT devices, such as Wi-Fi cameras and smart home appliances, connected to a home router alongside malicious entities. The dataset includes over 1 million records, with approximately 40073 normal instances and 585710 anomaly instances, offering a diverse set of network traffic data.
Primary categories (e.g., DoS, Mirai, MITM, Scan) and subcategories (e.g., MiraiUDP\_Flooding, DoSSynflooding, MiraiHostbruteforceg, MiraiAckflooding, MiraiHTTP\_Flooding, Scan\_Port\_OS, Scan\_Hostport, MITM\_ARP\_Spoofing). 
All these attributes enable the evaluation of machine learning models for intrusion detection.

\begin{table}[!t]
\centering
\caption{Dataset Discription}
\label{tab:table-2}
\resizebox{0.48\textwidth}{!}{%
\begin{tabular}{|c|c|c|c|c|}
\hline
\textbf{Dataset}       & \multicolumn{1}{l|}{\textbf{UNSW-NB15}} & \multicolumn{1}{l|}{\textbf{CIC-IDS2017}} & \multicolumn{1}{l|}{\textbf{IOTID20}} \\ \hline
\textbf{No. of Features} & 43 & 78 & 82 \\ \hline
\textbf{Training Samples} & 220862 & 245616 & 536385  \\ \hline
\textbf{Testing Samples} & 36811 & 40936 & 89398  \\ \hline
\end{tabular}%
}
\vspace{-2pt}
\end{table}

\subsection{Wasserstein CGAN-GP Based Oversampling}
The uneven distribution of various classes present in the datasets might strongly impact the performance of the classifier. 
Balancing the datasets is an essential first step in creating a machine learning \& deep learning prediction model.
To improve the dataset's balance, the study employs a method known as oversampling, which entails producing more samples for the minority classes. For such underrepresented classes, we used Wasserstein's Conditional Generative Adversarial Network with Gradient Penalty (WCGAN-GP) \cite{mckeever2020synthesising} method to balance and improve the accuracy of the learning model.


\begin{figure}[!t]
\centering
\includegraphics[width=0.5\textwidth, height=0.17\textheight]{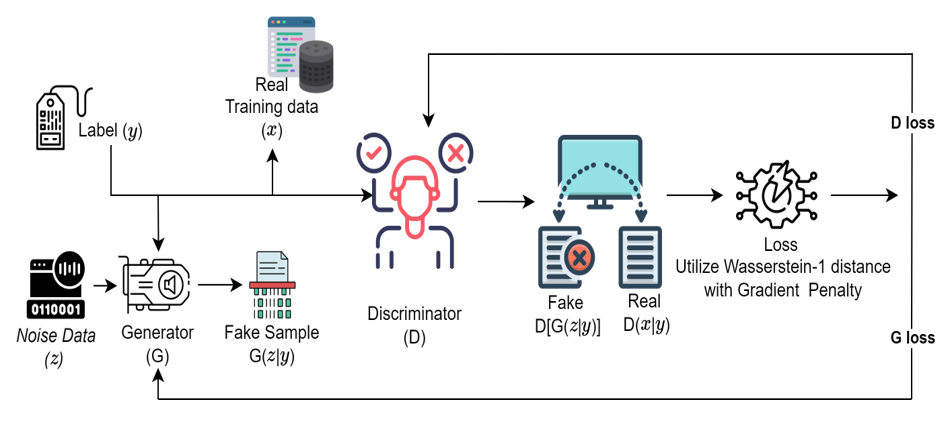}
\vspace{-12pt} 
\caption{Network Structure of WCGAN-GP.}
\vspace{-1pt} 
\label{fig:wgan}
\end{figure}


McKeever et al. \cite{mckeever2020synthesising} introduced the WCGAN-GP, a significant advancement in addressing the limitations of traditional GANs, particularly in the context of data augmentation for imbalanced datasets. By extending the WGAN-GP framework, the architecture of the WCGAN-GP model is shown in Figure \ref{fig:wgan}. The generator generates synthetic samples that mimic real data by receiving class labels and random noise vectors. On the other hand, two inputs are given to the discriminator network: one is the real training data samples from datasets, and the other is the synthetic samples generated from the generator network. Next, the discriminator attempts to discriminate between synthetic and real data samples. As the training progresses, the model's generator improves, eventually producing data that the discriminator finds difficult to distinguish from real samples. The gradient penalty helps ensure the model adheres to the Lipschitz continuity, effectively addressing issues such as mode collapse and vanishing gradients that are commonly encountered in traditional GANs. This makes WCGAN-GP particularly well-suited for enhancing data augmentation in imbalanced datasets, leading to better model performance, especially in scenarios where accurate minority attack detection is crucial. 
The objective function of WCGAN-GP is formulated as follows:
%
%
%

\begin{multline}
\small
               \min_G \max_D V(D, G) =  
 \mathbb{E}_{x \sim P_r} [D(x|y)] - \mathbb{E}_{\tilde{x} \sim P_g} [D(\tilde{x}|y)] \\ - \lambda \mathbb{E}_{\hat{x} \sim P_{\hat{x}}} \left[(\|\nabla_{\hat{x}} D(\hat{x}|y)\|_2 - 1)^2 \right]
\label{eq:Eq.(4)}
\end{multline}


Here, \(\lambda\) is the gradient penalty coefficient, and \(\hat{x}\) is sampled similarly to WGAN-GP. The loss functions for the discriminator and generator are defined as:

\textbf{Discriminator Loss:}

\begin{multline}
\small
L(D) =  -\mathbb{E}_{x \sim P_r} [D(x|y)] + \mathbb{E}_{\tilde{x} \sim P_g} [D(\tilde{x}|y)] \\
        + \lambda \mathbb{E}_{\hat{x} \sim P_{\hat{x}}} \left[(\|\nabla_{\hat{x}} D(\hat{x}|y)\|_2 - 1)^2 \right]
\label{eq:Eq.(5)}
\end{multline}

\textbf{Generator Loss:}

\begin{equation}
\small
L(G) = -\mathbb{E}_{\tilde{x} \sim P_g} [D(\tilde{x}|y)]
\label{eq:Eq.(generator)}
\end{equation}

Table \ref{table:parameter} shows the structural setting used by McKeever et al. \cite{mckeever2020synthesising} in their proposed WCGAN-GP model to optimize with certain values, e.g.,  no. of hidden layers is 3, a noise dimension of 256, and a feature count of 43 for the UNSW-NB15 dataset, 78 for the CIC-IDS2017 dataset and 82 for the IoTID20 dataset. In this paper, the architectural hyperparameter setting of the model was used as presented in Table \ref{table:hyperparameters}. After a certain number of epochs, the model's performance stabilizes and the critic loss converges to zero.

\begin{figure*}[!t]
\centering
\subfigure[{\scriptsize UNSW-NB15 Dataset}]{
    \includegraphics[width=0.65\columnwidth]{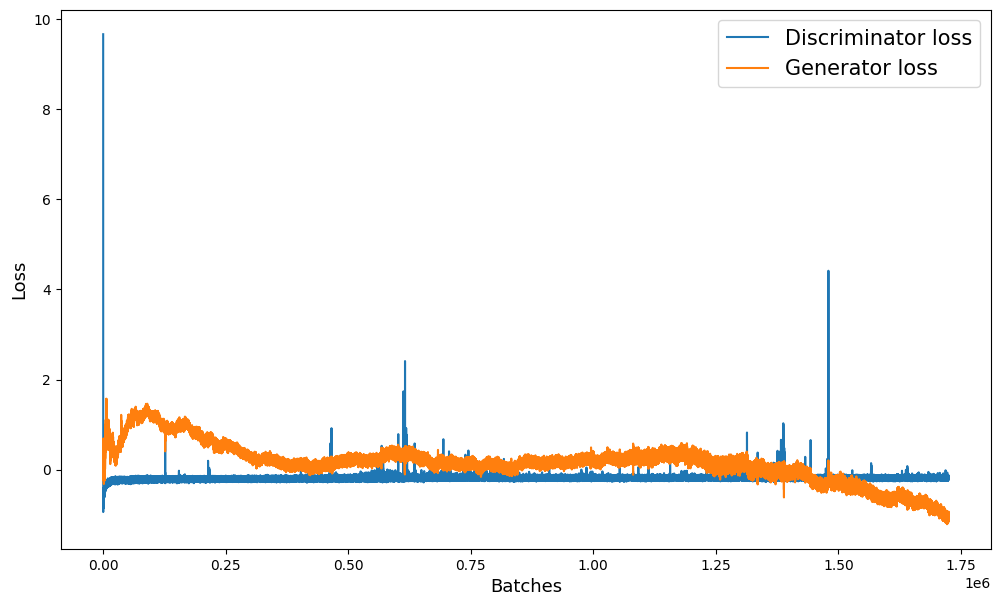}
    \label{fig:unsw_ganloss}
}
%
\subfigure[{\scriptsize CIC-IDS2017 Dataset}]{
    \includegraphics[width=0.65\columnwidth]{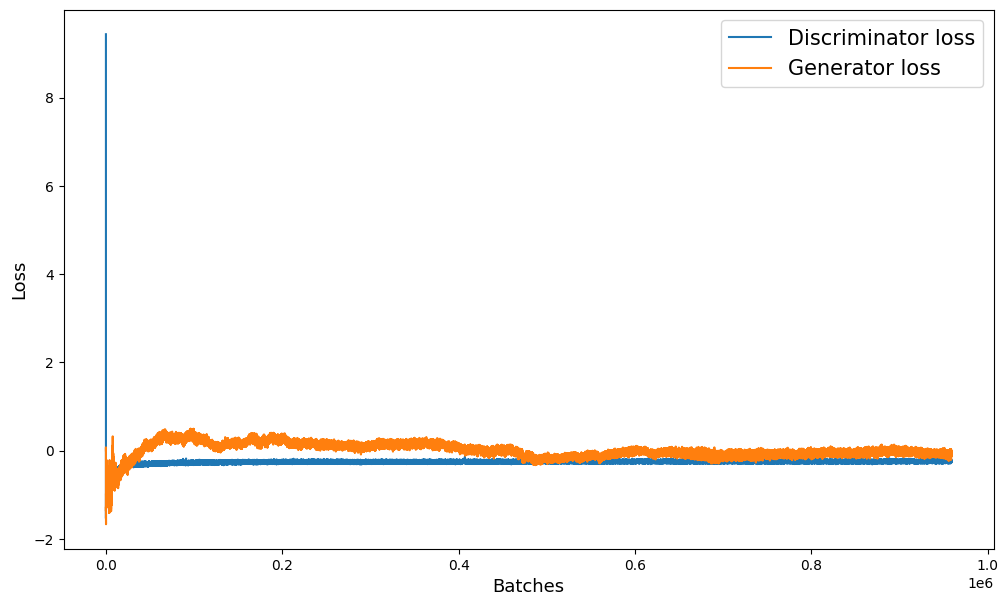}
    \label{fig:cic_ganloss}
}
%
\subfigure[{\scriptsize IOTID20 Dataset}]{
    \includegraphics[width=0.65\columnwidth]{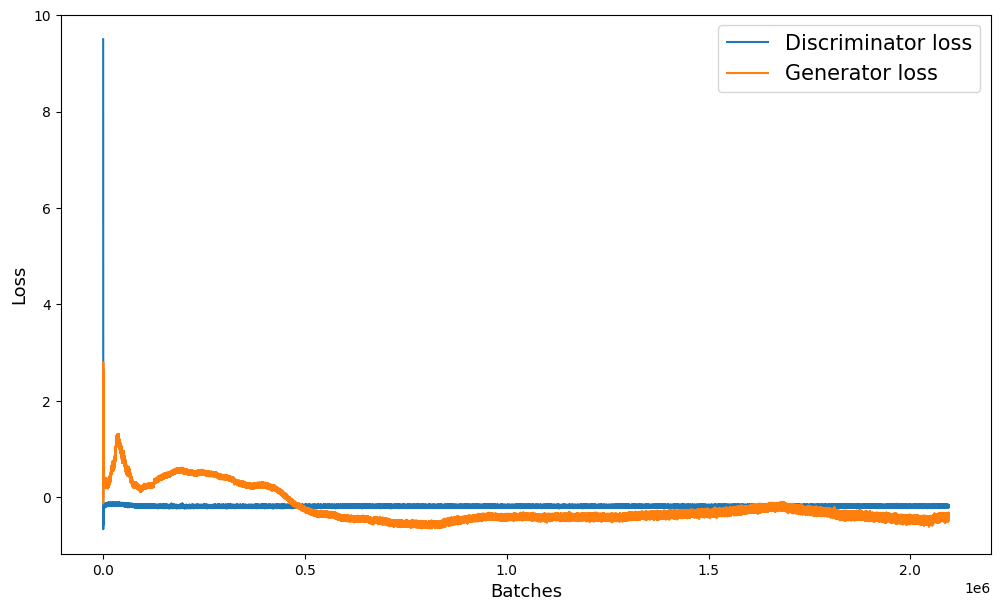}
    \label{fig:iotid20_ganloss}
}
\caption{WCGAN-GP loss curves demonstrate the training performance.}
\label{fig:gan_loss_comparison}
\end{figure*}

As shown in Figures \ref{fig:unsw_ganloss}, \ref{fig:cic_ganloss}, and \ref{fig:iotid20_ganloss}, the behavior of the model stabilizes after roughly 50 epochs, which shows the link between the discriminator loss, generator loss, and the number of batches. The loss increases quickly at first, but after some epochs, it converges toward zero. With the UNSW-NB15 dataset, a similar pattern is shown, with stability being attained after some epochs.


\begin{table*}[h!]
\centering
\caption{Structural Setting of the WCGAN-GP}
\label{table:parameter}
\renewcommand{\arraystretch}{1.3} 
\setlength{\tabcolsep}{6pt} 
\scriptsize 
\begin{tabular}{|l|c|c|c|c|c|c|}
\hline
\multirow{2}{*}{\textbf{Layers}} & \multicolumn{3}{c|}{\textbf{Generator}} & \multicolumn{3}{c|}{\textbf{Discriminator / Critic}} \\ \cline{2-7} 
                                 & \textbf{Size} & \textbf{Activation Function} & \textbf{Additional Info} & \textbf{Size} & \textbf{Activation Function} & \textbf{Additional Info} \\ \hline
\textbf{Input Layer}             & -             & -                          & Noise, label             & -             & -                          & Generated sample, label  \\ \hline
\textbf{Hidden Layer 1}          & 256           & Leaky ReLU                 & Dropout (rate=0.3)       & 1024          & Leaky ReLU                 & Dropout (rate=0.3)       \\ \hline
\textbf{Hidden Layer 2}          & 512           & Leaky ReLU                 & Dropout (rate=0.3)       & 512           & Leaky ReLU                 & Dropout (rate=0.3)       \\ \hline
\textbf{Hidden Layer 3}          & 1024          & Leaky ReLU                 & Dropout (rate=0.3)       & 256           & Leaky ReLU                 & Dropout (rate=0.3)       \\ \hline
\textbf{Output Layer}            & Data\_Dim     & Tanh                       & Generated data output    & 1             & Linear                     & Validity output          \\ \hline
\end{tabular}
\end{table*}


\begin{table}[h!]
\centering
\caption{Hyperparameter Values for WCGAN-GP}
\label{table:hyperparameters}
\renewcommand{\arraystretch}{1.2}
\setlength{\tabcolsep}{3pt}
\scriptsize 
\begin{tabular}{|p{3.0cm}|c|c|c|}
\hline
\textbf{Hyperparameters}                    & \multicolumn{3}{c|}{\textbf{Value in WCGAN-GP}} \\ \hline
                                            & \textbf{UNSW-NB15} & \textbf{CIC-IDS2017}  & \textbf{IOTID20}          \\ \hline
\textbf{No. of hidden layers}               & 3                          & 3                          & 3             \\ \hline
\textbf{Activation Function}                & \multicolumn{3}{c|}{ReLU / Tanh / Linear}                               \\ \hline
\textbf{Latent Dimension}                   & 64                         & 64                         & 64                                      \\ \hline
\textbf{Batch Size}                         & 128                        & 256                        & 256                                     \\ \hline
\textbf{Number of Discriminator Iterations} & 5                          & 5                          & 5                                       \\ \hline
\textbf{Gradient Penalty}                   & 10                         & 10                         & 10                                      \\ \hline
\textbf{Epochs}                             & 1000                       & 1000                       & 1000                                    \\ \hline
\textbf{Optimization Algorithm}             & \multicolumn{3}{c|}{Adam}                                                  \\ \hline
\textbf{Learning rate $(\alpha)$}                   & 0.0001                     & 0.0002                     & 0.0002                                  \\ \hline
\textbf{$\beta_1$}                                 & 0.02                       & 0.05                       & 0.05                                    \\ \hline
\textbf{$\beta_2$}                                 & 0.009                      & 0.9                      & 0.9                                     \\ \hline
\end{tabular}
\end{table}


\subsection{DualNetShield Algorithm}
DualNetShield is the most important component of our proposed framework. It is a two-phase algorithm to
enhance intrusion detection by addressing both major and minor attack classes. This proposed two-phase architecture is presented in Figure \ref{fig:dualnet}. Unlike single-phase systems, DualNetShield adopts a layered detection approach, ensuring precise classification even for underrepresented attacks.


\textbf{Phase 1: Major Attack Detection.}
The first phase of DualNetShield focuses on detecting major attacks, which are more common in datasets. It utilizes ten different machine learning models, each trained on an optimized set of features selected using Mutual Information Gain (MIG) \cite{battiti1994using}. MIG measures the dependency between features and the target variable, helping select the most informative features. 
This phase ensures efficient training and improves model accuracy by reducing irrelevant features. Machine learning models such as Support Vector Machine (SVM), Gradient Boosting (GB), Gaussian Naive Bayes, Decision Tree, Bagging Classifier, AdaBoost Classifier, Extra Tree Classifier, MLP Classifier, Logistic Regression (LR), and XGBoost are employed to detect major attacks. If an instance is identified as a major attack, it is immediately classified. Minor or normal instances proceed to the second phase.

\textbf{Phase 2: Minor Attack Detection.}
The second phase of DualNetShield refines the classification of instances identified as minor attacks or normal traffic. It addresses the imbalance of minority attack classes with advanced techniques:
\emph{1) Refined Feature Selection:} Mutual Information Gain is reapplied to select features specifically suited for detecting rare attack types, enhancing model precision.
Only the most relevant features are retained to improve accuracy in identifying minority attacks.
\emph{2) Handling Imbalanced Data Through Random Downsampling:} To balance the dataset, the majority class is downsampled randomly. This prevents bias towards the majority class, allowing the model to effectively detect underrepresented attack types.
\emph{3) Using Random Forest for Minor Attack Detection:} Random Forest is chosen for its ability to handle imbalanced datasets and its ensemble nature, which reduces variance and improves robustness.
The model’s feature importance capabilities provide insights into subtle patterns of minor attacks and normal traffic, enhancing its detection accuracy.

This two-phase architecture enables DualNetShield to provide precise detection across both major and minor attacks, creating a comprehensive intrusion detection system with high accuracy. Details of the DualNetShield algorithm are provided in Algorithm \ref{alg:dualnet_shield}.


\begin{algorithm}[!t]
\small
\caption{DualNetShield Algorithm}
\label{alg:dualnet_shield}
\KwIn{New training data $D_{train} = \{f_1,f_2,\dots,f_n\}$, and Original test data $D_{test}$}
\KwOut{Classification result $y \in \{\text{Attack}, \text{Normal}\}$, Best model combination}

\textbf{Procedure:} DualNetShield\;

\While{DualNetShield is active}{
    \tcp{Phase 1: Major Attack Detection (Group 1 with Multiple Models)}
    Process $D_{train}$ to create Group 1 (All Data)\;
    Extract features from Group 1 using Mutual Information Gain\;
    
    \For {Each model $M_i$ in $\{M_1, M_2, \dots, M_n\}$}{
        Train $M_i$ on Group 1\;
        \If{$M_i$ detects a Major Attack}{
            Store the result of $M_i$ as $y_i \gets \text{Attack}$\;
        }
        \Else{
            Store the result of $M_i$ as $y_i \gets \text{Proceed to Phase 2}$\;
        }
    }
    
    \tcp{Phase 2: Minor Attack Detection (Group 2 with Random Forest)}
    Process $D_{train}$ to create Group 2 (Minority Attack)\;
    Extract features from Group 2 using Mutual Information Gain\;
    
    \If{Class imbalance is detected in Group 2}{
        Downsample frequent instances in Group 2\;
    }
    
    Train the Random Forest model $RF$ on the processed Group 2 data\;
    
    \For {Each stored result $y_i$ from Group 1}{
        \If{$y_i$ == \text{Proceed to Phase 2}}{
            \If{$RF$ detects a Minor Attack}{
                Store the combination result of $M_i$ and $RF$ as $y_i \gets \text{Attack}$\;
            }
            \Else{
                Store the combination result of $M_i$ and $RF$ as $y_i \gets \text{Normal}$\;
            }
        }
    }
    
    \tcp{Final Selection: Best Model Combination}
    Evaluate all combinations of Group 1 models and Group 2 Random Forest based on performance metrics (e.g., accuracy, precision, recall)\;
    Select the best-performing combination model $\{M_{\text{best}}, RF\}$\;
    
    \textbf{Return} $y$ from the best combination model, Best model combination $\{M_{\text{best}}, RF\}$\;
}
\end{algorithm}




\section{Experiment and Evaluation } 
\label{sec:experiment}
The experiments are conducted on an ASUS ROG STRIX B760-F workstation with a 13th Gen Intel Core i9-13900K with 24 cores @ 5.80 GHz, 64 GB of memory, and an NVIDIA graphics card with a resolution of 3840$\times$1080. The system operates on Ubuntu 22.04.5 LTS, and the performance of AI models is evaluated using TensorFlow 2.13.1, Matplotlib 3.7.5, and scikit-learn 1.2.2.

\subsection{Data Pre-processing}

Before training a model, we carry out a thorough \textbf{data cleaning} procedure that comprises the following steps to guarantee data quality and reliability. \par


\begin{enumerate}
    \item Handling missing values: Depending on the degree of missing data, either eliminate incomplete entries or use imputation techniques to identify and rectify missing data.
    \item Outlier removal: is the process of employing statistical methods or domain-specific expertise to identify and control outliers that have the potential to distort results and poor model performance.
    \item Removing or fixing null and infinite values: To avoid problems during model training and evaluation, remove or fix null and infinite values.
\end{enumerate}

After cleaning the data, we prepare the dataset for model training by applying the following \textbf{data transformation} steps:

\begin{enumerate}
    \item Label encoding: Using label encoding, which ensures compatibility with ML/DL algorithms by assigning an integer to each category starting at zero, categorical features, including the target variable, are converted into numerical values.
    \item Feature scaling: It is the process of putting features on a comparable scale by using both standardization and normalization.
    Standardization is the process of rescaling features to have a mean of zero and a standard deviation of one:
   \[
    X_{standardized} = \frac{X - \mu}{\sigma}
     \]
    where \(X\) is the original value, \(\mu\) is the mean, and \(\sigma\) is the standard deviation.
While normalization involves scaling features to a unit norm using the L2 norm:
   \[
    X_{normalized} = \frac{X}{\|X\|}
    \]
    where \(X\) is the original data point, and \(\|X\|\) is its Euclidean norm.
\end{enumerate}

After data transformation, we performed the \textbf{data splitting}. As illustrated in Table \ref{tab:table-2}, the dataset is scaled and then divided into training and testing sets in a 6:1 ratio to preserve a balance between training the model and impartially assessing its performance.

\begin{table*}
\caption{Number of samples across different datasets and attack classes}
\centering
\label{tab:dataset_table}
\scriptsize
\renewcommand{\arraystretch}{1.2}
\setlength{\tabcolsep}{3pt}
\begin{tabular}{|l|l|p{3.44cm}|p{2.0cm}|r|r|r|r|}
\hline
\multirow{2}{*}{\textbf{Datasets}} & \multirow{2}{*}{\textbf{Attack Type}} & \multirow{2}{*}{\textbf{Classes}} & \multirow{2}{*}{\textbf{Original Samples}} & \multicolumn{2}{c|}{\textbf{Original Samples}} & \multicolumn{2}{c|}{\textbf{New Training Samples}} \\ \cline{5-8}
& & & & \textbf{Training} & \textbf{Testing} & \textbf{Synthetic} & \textbf{Augmented} \\ \hline

\multirow{19}{*}{\rotatebox{90}{\textbf{CIC-IDS2017}}} & 
\multirow{1}{*}{\rotatebox{0}{Non--attack}} & BENIGN              & 227,199 & 194,742 & 32,457 & -- & 194,742 \\ \cline{2-8}
& \multirow{3}{*}{\rotatebox{0}{Major Attack}} & DoS Hulk            & 23,239 & 19,919 & 3,320 & -- & 19,919 \\ \cline{3-8}
& & DDoS                & 12,668 & 10,858 & 1,810 & -- & 10,858 \\ \cline{3-8}
& & PortScan            & 15,780 & 13,526 & 2,254 & -- & 13,526 \\ \cline{2-8}

& \multirow{11}{*}{\rotatebox{0}{Minor Attack}} & \textbf{Bot}                 & 1,956 & \textcolor{red}{1,677} & 279 & \textbf{4,000} & \textcolor{red}{\textbf{5,677}} \\ \cline{3-8}
& & \textbf{DoS GoldenEye}       & 996 & \textcolor{red}{854} & 142 & \textbf{4,000} & \textcolor{red}{\textbf{4,854}} \\ \cline{3-8}
& & \textbf{DoS Slowhttptest}    & 561 & \textcolor{red}{481} & 80 & \textbf{4,000} & \textcolor{red}{\textbf{4,481}} \\ \cline{3-8}
& & \textbf{DoS Slowloris}       & 613 & \textcolor{red}{525} & 88 & \textbf{4,000} & \textcolor{red}{\textbf{4,525}} \\ \cline{3-8}
& & \textbf{FTP Patator}         & 766 & \textcolor{red}{656} & 110 & \textbf{4,000} & \textcolor{red}{\textbf{4,656}} \\ \cline{3-8}
& & \textbf{Heartbleed}          & 11 & \textcolor{red}{9} & 2 & \textbf{4,000} & \textcolor{red}{\textbf{4,009}} \\ \cline{3-8}
& & \textbf{Infiltration}        & 36 & \textcolor{red}{31} & 5 & \textbf{4,000} & \textcolor{red}{\textbf{4,031}} \\ \cline{3-8}
& & \textbf{SSH Patator}         & 547 & \textcolor{red}{469} & 78 & \textbf{4,000} & \textcolor{red}{\textbf{4,469}} \\ \cline{3-8}
& & \textbf{Web Attack - Brute Force} & 1,507 & \textcolor{red}{1,292} & 215 & \textbf{4,000} & \textcolor{red}{\textbf{5,292}} \\ \cline{3-8}
& & \textbf{Web Attack - SQL Injection} & 21 & \textcolor{red}{18} & 3 & \textbf{4,000} & \textcolor{red}{\textbf{4,018}} \\ \cline{3-8}
& & \textbf{Web Attack - XSS}    & 652 & \textcolor{red}{559} & 93 & \textbf{4,000} & \textcolor{red}{\textbf{4,559}} \\ \hline

\multirow{9}{*}{\rotatebox{90}{\textbf{UNSW-NB15}}} & 
\multirow{1}{*}{\rotatebox{0}{Non--attack}} & Normal              & 93,000 & 79,673 & 13,327 & -- & 79,673 \\ \cline{2-8}
& \multirow{5}{*}{\rotatebox{0}{Major Attack}} & Generic             & 58,871 & 50,488 & 8,383 & -- & 50,488 \\ \cline{3-8}
& & Exploits            & 44,525 & 38,044 & 6,481 & -- & 38,044 \\ \cline{3-8}
& & Fuzzers             & 24,246 & 20,830 & 3,416 & -- & 20,830 \\ \cline{3-8}
& & DoS                 & 16,353 & 14,077 & 2,276 & -- & 14,077 \\ \cline{3-8}
& & Reconnaissance                & 13,987 & 11,984 & 2,003 & -- & 11,984 \\ \cline{2-8}

& \multirow{4}{*}{\rotatebox{0}{Minor Attack}}  & \textbf{Analysis}            & 2,677 & \textcolor{red}{2,292} & 385 & \textbf{4,000} & \textcolor{red}{\textbf{6,292}} \\ \cline{3-8}

& & \textbf{Backdoor}            & 2,329 & \textcolor{red}{2,017} & 312 & \textbf{4,000} & \textcolor{red}{\textbf{6,017}} \\ \cline{3-8}
& & \textbf{Shellcode}           & 1,511 & \textcolor{red}{1,309} & 202 & \textbf{4,000} & \textcolor{red}{\textbf{5,309}} \\ \cline{3-8}
& & \textbf{Worms}               & 174 & \textcolor{red}{148} & 26 & \textbf{4,000} & \textcolor{red}{\textbf{4,148}} \\ \hline

\multirow{9}{*}{\rotatebox{90}{\textbf{IOTID20}}} & 
\multirow{1}{*}{\rotatebox{0}{Non--attack}} & Normal              & 40,073 & 34,321 & 5,752 & -- & 34,321 \\ \cline{2-8}
& \multirow{6}{*}{\rotatebox{0}{Major Attack}} & MiraiUDP\_Flooding             & 183,554 & 157,384 & 26,170 & -- & 157,384 \\ \cline{3-8}
& & DoSSynflooding             & 59,391 & 51,003 & 8,388 & -- & 51,003 \\ \cline{3-8}
& & MiraiHostbruteforceg             & 121,181 & 103,892 & 17,289 & -- & 103,892 \\ \cline{3-8}
& & MiraiAckflooding            & 55,124 & 47,211 & 7,913 & -- &  55,124 \\ \cline{3-8}
& & MiraiHTTP\_Flooding           & 55,818 & 47,769 & 8,049 & -- & 55,818 \\ \cline{3-8}
& & Scan\_Port\_OS            & 53,073 & 45,477 & 7,596 & -- & 53,073 \\ \cline{2-8}

 & \multirow{2}{*}{\rotatebox{0}{Minor Attack}} & \textbf{Scan\_Hostport}            & 22,192 & \textcolor{red}{19,022} & 3,170 & \textbf{12,000} & \textcolor{red}{\textbf{31,022}} \\ \cline{3-8}
& & \textbf{MITM\_ARP\_Spoofing}           & 35,377 & \textcolor{red}{30,306} & 5,071 & \textbf{12,000} & \textcolor{red}{\textbf{42,306}} \\ \hline

\end{tabular}
\end{table*}


\subsection{Training of WCGAN-GP and Hyperparameters Tuning}
After data preprocessing, the next step involved the training of the WCGAN-GP to address the class imbalance in the UNSW-NB15, CIC-IDS2017, and IoTID20 datasets. The training process alternated between updating the generator (G) and discriminator (D) to achieve stable convergence and generate synthetic samples for minority attack classes. \\

The generator \( G \) takes in a random noise vector \( z \), which follows a Gaussian distribution, and a category label \( y \) as inputs. The generator then produces new attack samples, \( G(z) \), but initially, these samples have low similarity to real ones. Next, we fix the generator \( G \) and train the discriminator \( D \). During this phase, real and generated samples are mixed as input to the discriminator, which outputs the probability of each sample being real or generated. These probability values are then converted into predicted labels using an activation function. The objective function of the discriminator is shown in Equation \ref{eq:Eq.(4)}.

Once the discriminator is trained, we move on to train the generator \( G \) through the combined process of \( G - D \). After training \( D \), its ability to distinguish real from fake samples improves. At this stage, the generator learns to create more realistic samples. 
This process of alternately training the generator and discriminator continues until the loss reaches a preset threshold or the specified number of training cycles is completed. The Adam optimizer is used to update the gradient penalty and improve the overall loss function.
For the WCGAN-GP, the learning rate \( \alpha \), the gradient penalty \( \lambda \), and the generator and discriminator are structured with hidden layers sized, batch size, and the model undergoes training epochs are shown in Table \ref{table:parameter} and \ref{table:hyperparameters}.


\subsection{Building New Training Datasets}
After training, WCGAN-GP is used to create new attack samples for minority categories as shown in Table \ref{tab:dataset_table}, based on the number of attack samples in the original dataset. These generated attack samples are then combined with the original training data to form a new, more balanced dataset. This helps to address the issue of imbalanced data and increases the variety of samples for training.

There are 15 classes in the CIC-IDS2017 dataset: four are majority classes (DoSHulk, DDoS, PortScan, and BENIGN) and the rest are minority classes. There are 10 classes in the UNSW-NB15 dataset, with six (Normal, Generic, Exploits, Fuzzers, Dos, and Reconnaissance) making up the bulk (majority) and the other four are in the minority. After applying the WCGAN-GP method, we generated 4,000 samples for each minority class in both datasets. These synthetic samples were then added to the real training data to create a new training dataset. However, in the IOTID20 dataset, by applying the WCGAN-GP method, we generated 12,000 samples for the minority classes. Following the application of the WCGAN-GP model, Table \ref{tab:dataset_table} displays the distribution of data across several classes in the training datasets.


\begin{figure*}[!t]
\centering
\fontsize{14}{16}\selectfont 
\includegraphics[width=0.80\textwidth]{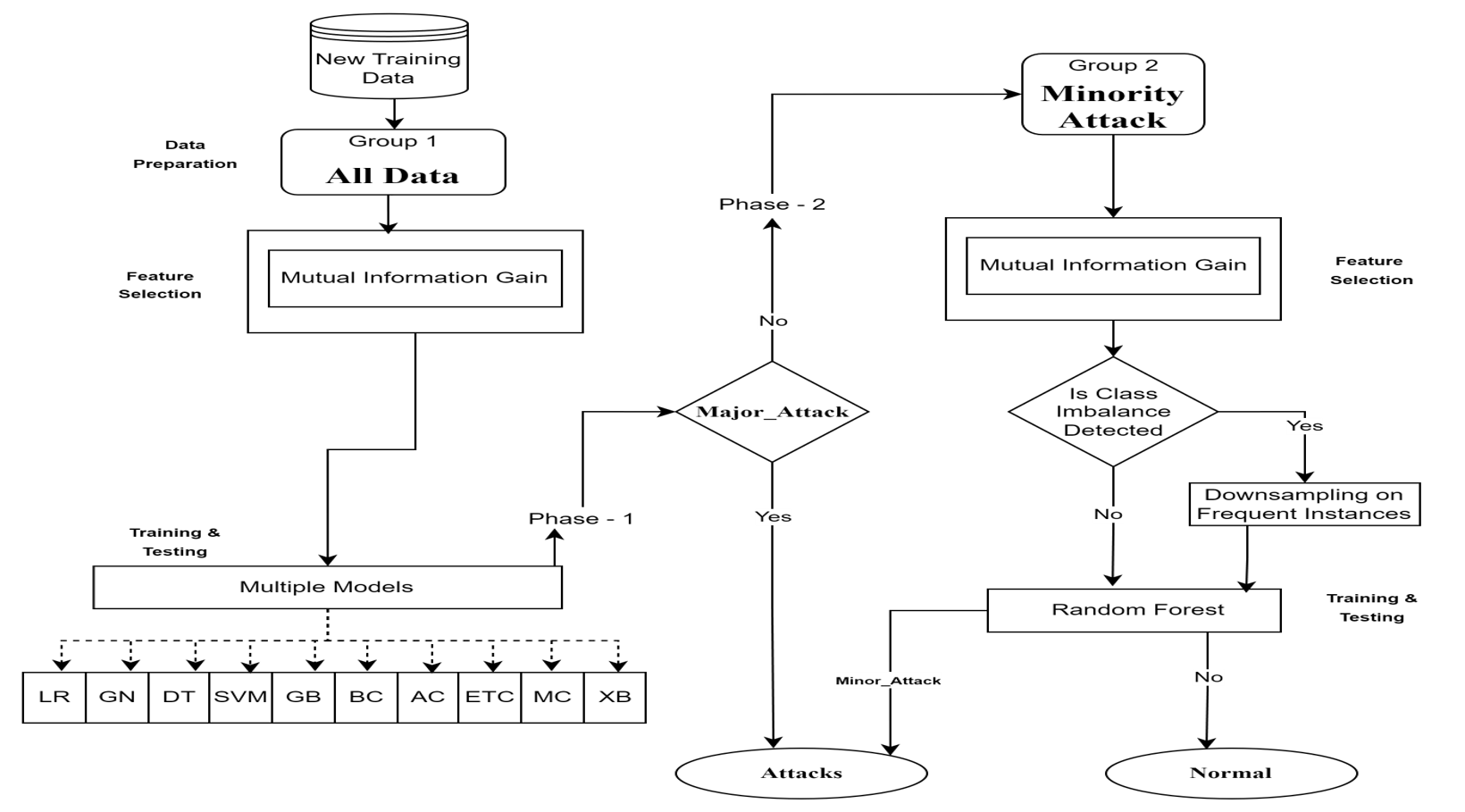}
\caption{Workflow diagram of DualNetSheild algorithm.}
\vspace{-4pt} 
\label{fig:dualnet}
\end{figure*}


\subsection{Detection of Attacks}

After generating synthetic data using the WCGAN-GP model, the synthetic and real datasets are combined to create a new training set. This new dataset is then processed through the DualNetShield framework Figure \ref{fig:dualnet}, which operates in two distinct phases. 

In \textbf{Phase 1}, the entire dataset undergoes feature selection using Mutual Information, and the top 30 features related to major attacks are selected. These features are subsequently used to train a variety of machine learning models, including Logistic Regression, Gaussian Naive Bayes, Decision Tree, Support Vector Machine, Gradient Boosting, Bagging Classifier, AdaBoost Classifier, Extra Tree Classifier, MLP Classifier, and XGBoost. The purpose of this phase is to detect major attacks, which are more common and typically easier to identify. If a major attack is detected, the system classifies the event as an attack. However, if the analysis suggests the presence of a minor attack or normal traffic, the process advances to Phase 2.


\begin{table}[!t]
\centering
\caption{Model combinations for DualNetShield}
\label{table:reference_models}
\begin{tabular}{|l|c|}
\hline
\textbf{Reference} & \textbf{Model Name} \\ \hline
Logistic Regression + Random Forest & M1 \\ \hline
Gaussian NB + Random Forest & M2 \\ \hline
Decision Tree + Random Forest & M3 \\ \hline
Support Vector Machine + Random Forest & M4 \\ \hline
Gradient Boosting + Random Forest & M5 \\ \hline
Bagging Classifier + Random Forest & M6 \\ \hline
AdaBoost Classifier + Random Forest & M7 \\ \hline
Extra Tree Classifier + Random Forest & M8 \\ \hline
MLP Classifier + Random Forest & M9 \\ \hline
XGBoost + Random Forest & M10 \\ \hline
\end{tabular}
\vspace{0.3cm}  
\end{table}

In \textbf{Phase 2}, the focus shifts to identifying minority attacks, which are less frequent but often more difficult to detect. Once again, Mutual Information is applied to select the top 30 features relevant to these minority attacks. To address any class imbalance that may exist in the dataset—where normal traffic significantly outnumbers attack samples—a random undersampling technique is employed to ensure a balanced representation. If no significant imbalance is detected, the process proceeds without undersampling. The selected features are then used to train a Random Forest model, which is responsible for classifying the data as either a minor attack or normal traffic. This two-phase approach ensures that both major and minor attacks are effectively identified.

To evaluate the whole performance of the proposed system, the results from WCGAN-GP and each machine learning model from Phase 1 are paired with the Random Forest model from Phase 2 in DualNetShield, creating a HybridGuard detection model. These DualNetShield models are named as follows in Table~\ref{table:reference_models}, showcasing their ability to improve both major \& minor attack detection in imbalanced datasets.


\begin{table}[!t]
\centering
\caption{Evaluation factors}
\label{tab:metric}
\begin{tabular}{|l|c|c|}
\hline
\textbf{Indicator}   & \textbf{Actual Label} & \textbf{Prediction} \\ \hline
True Positive ($TP$)  & Attack          & Attack              \\ \hline
False Positive ($FP$) & Normal          & Attack              \\ \hline
True Negative ($TN$)  & Normal          & Normal              \\ \hline
False Negative ($FN$) & Attack          & Normal              \\ \hline
\end{tabular}
\vspace{-2pt}
\end{table}

\subsection{Evaluation Metrics}
To evaluate the performance of the HybridGuard, six common metrics are used: accuracy, F1 score, precision, recall, false alarm rate (FAR), and confusion matrix.
A confusion matrix is a tabular representation that displays the number of correct and incorrect predictions for each class. It provides valuable insight into the model’s performance by categorizing predictions into True Positives, True Negatives, False Positives, and False Negatives, helping to assess classification accuracy and errors. Therefore, the effectiveness of an IDS can be evaluated using these four key factors, as illustrated in Table~\ref{tab:metric}.
Accuracy represents the ratio of correctly classified samples to the total number of samples evaluated, as in Equation~\eqref{eq:Eq.(accuracy)}.
Precision is the proportion of true positive predictions relative to the total predicted positive instances, as in Equation~\eqref{eq:Eq.(precision)}.
Recall is the proportion of correctly identified positive instances out of all actual positive cases, as in Equation~\eqref{eq:Eq.(recall)}.
The F1 score represents the harmonic mean of precision and recall, considering both false positives and false negatives as in Equation~\eqref{eq:Eq.(f1)}. When false positives and false negatives are unevenly distributed, the F1 score provides a more balanced assessment of the model’s performance.
Equation~\eqref{eq:Eq.(far)} shows the calculation of FAR, which refers to the percentage of normal instances mistakenly classified as attacks. This metric is crucial as it indicates how frequently the system generates unnecessary alerts, potentially leading to overestimation and increased need for human intervention.

\begin{equation}
 Accuracy = \frac{TP + TN}{TP + TN + FP + FN}
\label{eq:Eq.(accuracy)}
\end{equation}
%

\begin{equation}
Precision,~~ P= \frac{TP}{(TP+FP)}
\label{eq:Eq.(precision)}
\end{equation}

\begin{equation}
Recall,~~ R = \frac{TP}{(TP+FN)} 
\label{eq:Eq.(recall)}
\end{equation}

\begin{equation}
F1~score = \frac{2*P*R}{(P+R)}
\label{eq:Eq.(f1)}
\end{equation}

%
\begin{equation}
FAR = \frac{FP}{FP + TN}
\label{eq:Eq.(far)}
\end{equation}


Since our dataset includes a significant amount of benign data, using a weighted average F1 score does not offer a comprehensive evaluation. Instead, we utilize the macro-average F1 score, which calculates the metric separately for each label and then averages the results, ensuring equal consideration for all labels.

\section{Results and Discussion } \label{sec:result}

\subsection{Performance Based on DualNetShield}
In this section, we discuss the performance of DualNetShield using different combinations of classifiers. DualNetShield, as the foundational model, employs a variety of classifiers such as SVM, Gradient Boosting, etc. with Random Forest to detect major and minor attacks.

For the UNSW-NB15 dataset as shown in Table \ref{tab:dualnet_model_compare}, \textbf{Model M4} emerged as the best performer, achieving the highest scores: Accuracy (94.38\%), F1 Score (95.46\%), Precision (97.45\%), Recall (94.38\%), and a False Alarm Rate (FAR) of 9.22\%. This made Model M4 the most effective for detecting both attack and normal traffic in this dataset. Similarly, for the CIC-IDS2017 dataset, \textbf{Model M2} demonstrated the best results with an Accuracy of 78.89\%, F1 Score of 86.36\%, Precision of 98.01\%, Recall of 78.89\%, and a FAR of 26.09\%. In the case of the IoTID20 dataset, \textbf{Model M10} stood out as the most effective. It achieved an Accuracy of 99.6\%, F1 Score of 99.6\%, Precision of 99.6\%, Recall of 99.6\%, and a FAR of 0.6\%.

This initial evaluation of DualNetShield models helped in identifying the most effective model combinations for each dataset. 
However, despite these results, DualNetShield faced challenges with data imbalance and struggled to effectively detect minor attacks. These issues resulted in higher false alarm rates and lower performance in handling rare attack types.

\begin{table*}[t]
\caption{Performance of DualNetShield Models}
\centering
\label{tab:dualnet_model_compare}
\resizebox{0.99 \textwidth}{!}{%
\small 
\begin{tabular}{|c|c c c c c|c c c c c|c c c c c|}
\hline
\multirow{2}{*}{\textbf{Models (DualNetShield)}} & 
\multicolumn{5}{c|}{\textbf{UNSW-NB15 Dataset}} & 
\multicolumn{5}{c|}{\textbf{CIC-ID2017 Dataset}} &
\multicolumn{5}{c|}{\textbf{IOTID20 Dataset}} \\ \cline{2-16} 
 & \textbf{Accuracy} & \textbf{F1-Score} & \textbf{Precision} & \textbf{Recall} & \textbf{FAR} & \textbf{Accuracy} & \textbf{F1-Score} & \textbf{Precision} & \textbf{Recall} & \textbf{FAR} & 
 \textbf{Accuracy} & \textbf{F1-Score} & \textbf{Precision} & \textbf{Recall} & \textbf{FAR} \\ \hline
M1 & 94.25 & 95.29 & 97.17 & 94.25 & 5.75 & 58.74 & 71.67 & 97.42 & 58.74 & 46.19 & 97.1 & 98.0 & 98.4 & 97.9 & 0.1   \\ 

\textcolor{red}{\textbf{M2}} & 76.49 & 80.46 & 92.44 & 76.49 & 11.94 & \textcolor{red}{\textbf{78.89}} & \textcolor{red}{\textbf{86.36}} & \textcolor{red}{\textbf{98.01}} & \textcolor{red}{\textbf{78.89}} & \textcolor{red}{\textbf{26.09}} & 75.2 & 74.9 & 92.2 & 75.2 & 0.8 \\ 

M3 & 91.46 & 93.37 & 96.75 & 91.46 & 12.99 & 61.15 & 72.75 & 97.53 & 61.15 & 47.94 & 99.2 & 99.2 & 99.3 & 99.2 & 0.8 \\ 

\textcolor{red}{\textbf{M4}} & \textcolor{red}{\textbf{94.38}} & \textcolor{red}{\textbf{95.46}} & \textcolor{red}{\textbf{97.45}} & \textcolor{red}{\textbf{94.38}} & \textcolor{red}{\textbf{9.22}} & 58.75 & 71.50 & 97.45 & 58.75 & 47.01 & 98.4 & 98.5 & 98.7 & 98.4 & 0.2\\ 

M5 & 93.59 & 94.74 & 96.97 & 93.59 & 10.02 & 60.62 & 72.54 & 97.43 & 60.62 & 47.62 & 99.4 & 99.4 & 99.5 & 99.4 & 0.2 \\ 

M6 & 91.29 & 93.31 & 96.85 & 91.29 & 14.03 & 61.11 & 72.72 & 97.52 & 61.11 & 47.98 & 99.3 & 99.3 & 99.3 & 99.3 & 0.8 \\ 

M7 & 93.65 & 94.63 & 96.65 & 93.65 & 7.29 & 60.46 & 72.60 & 97.45 & 60.46 & 47.01 & 98.6 & 98.6 & 98.8 & 98.6 & 0.2\\ 

M8 & 92.23 & 94.00 & 97.12 & 92.23 & 13.08 & 61.14 & 72.77 & 97.53 & 61.14 & 47.84 & 99.1 & 99.1 & 99.2 & 99.1 & 0.7 \\ 

M9 & 91.74 & 93.48 & 96.74 & 91.74 & 13.06 & 62.61 & 74.07 & 97.51 & 62.61 & 45.85 & 98.6 & 98.6 & 98.8 & 98.6 & 0.5 \\ 

\textcolor{red}{\textbf{M10}} & 92.08 & 93.83 & 96.96 & 92.08 & 13.09 & 61.27 & 72.84 & 97.54 & 61.27 & 47.82 & \textcolor{red}{\textbf{99.6}} & \textcolor{red}{\textbf{99.6}} & \textcolor{red}{\textbf{99.6}} & \textcolor{red}{\textbf{99.6}} & \textcolor{red}{\textbf{0.6}} \\ \hline

\end{tabular}%
}
\end{table*}

\begin{table*}[t]
\caption{Performance of HybridGuard Models}
\centering
\label{tab:table_hybridguard_model}
\resizebox{0.99 \textwidth}{!}{%
\small 
\begin{tabular}{|c|c c c c c|c c c c c|c c c c c|}
\hline
\multirow{2}{*}{\textbf{Models (HybriGaurds)}} & 
\multicolumn{5}{c|}{\textbf{UNSW-NB15 Dataset}} & 
\multicolumn{5}{c|}{\textbf{CIC-ID2017 Dataset}} &
\multicolumn{5}{c|}{\textbf{IOTID20 Dataset}} \\ \cline{2-16} 
 & \textbf{Accuracy} & \textbf{F1-Score} & \textbf{Precision} & \textbf{Recall} & \textbf{FAR} & \textbf{Accuracy} & \textbf{F1-Score} & \textbf{Precision} & \textbf{Recall} & \textbf{FAR} &
 \textbf{Accuracy} & \textbf{F1-Score} & \textbf{Precision} & \textbf{Recall} & \textbf{FAR} \\ \hline
 
WCGAN-GP + M1 & 94.47 & 94.67 & 95.72 & 94.47 & 1.27 & 91.33 & 90.45 & 92.26 & 91.33 & 1.20 & 96.2 & 96.5 & 97.6 & 96.2 & 0.2 \\ 

\textcolor{red}{\textbf{WCGAN-GP + M2}} & 79.38 & 80.18 & 89.56 & 79.38 & 3.81 & \textcolor{red}{\textbf{99.87}} & \textcolor{red}{\textbf{99.87}} & \textcolor{red}{\textbf{99.89}} & \textcolor{red}{\textbf{99.87}} & \textcolor{red}{\textbf{0.14}} & 74.4 & 73.7 & 91.7 & 74.4 & 0.7 \\ 

WCGAN-GP + M3 & 94.74 & 94.96 & 96.09 & 94.74 & 3.31 & 98.64 & 98.77 & 99.02 & 98.64 & 1.25 & 99.2 & 99.2 & 99.3 & 99.2 & 0.7 \\ 

\textcolor{red}{\textbf{WCGAN-GP + M4}} & \textcolor{red}{\textbf{97.07}} & \textcolor{red}{\textbf{97.09}} & \textcolor{red}{\textbf{97.47}} & \textcolor{red}{\textbf{97.07}} & \textcolor{red}{\textbf{2.00}} & 95.44 & 95.34 & 95.92 & 95.44 & 1.25 & 98.2 & 98.3 & 98.5 & 98.2 & 0.1 \\ 

WCGAN-GP + M5 & 96.71 & 96.77 & 96.71 & 96.71 & 2.05 & 97.73 & 97.85 & 98.12 & 97.73 & 1.25 & 99.4 & 99.4 & 99.4 & 99.4 & 0.1 \\ 

WCGAN-GP + M6 & 94.90 & 95.06 & 96.12 & 94.90 & 3.59 & 98.64 & 98.77 & 99.02 & 98.64 & 1.25 & 99.3 & 99.3 & 99.3 & 99.3 & 0.7 \\ 

WCGAN-GP + M7 & 96.15 & 96.31 & 96.56 & 96.15 & 1.75 & 96.89 & 96.98 & 97.3 & 96.89 & 1.25 & 98.0 & 98.2 & 98.3 & 98.0 & 0.2 \\ 

WCGAN-GP + M8 & 95.89 & 95.79 & 95.94 & 95.89 & 3.14 & 98.75 & 98.85 & 98.94 & 98.75 & 1.25 & 99.1 & 99.1 & 99.2 & 99.1 & 0.5 \\ 

\textcolor{red}{\textbf{WCGAN-GP + M9}} & 95.87 & 95.79 & 96.09 & 95.87 & 2.90 & 97.75 & 97.85 & 98.12 & 97.75 & 1.25 & \textcolor{red}{\textbf{99.6}} & \textcolor{red}{\textbf{99.6}} & \textcolor{red}{\textbf{99.6}} & \textcolor{red}{\textbf{99.6}} & \textcolor{red}{\textbf{0.2}} \\ 

WCGAN-GP + M10 & 79.38 & 80.18 & 89.56 & 79.38 & 3.81 & 98.69 & 98.82 & 99.02 & 98.69 & 1.25 & 99.6 & 99.6 & 99.6 & 99.6 & 0.5 \\ \hline
\end{tabular}%
}
\end{table*}

\begin{figure*}[!t]
\centering
\subfigure[{\scriptsize UNSW-NB15 Dataset}]{
    \includegraphics[width=0.65\columnwidth]{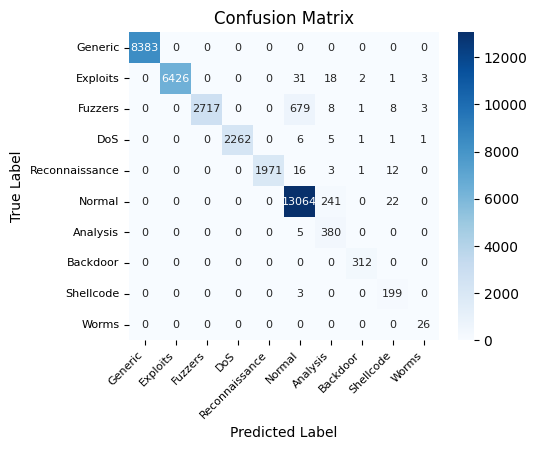}
    \label{fig:unsw-nb15_confusion}
}
%
\subfigure[{\scriptsize CIC-IDS2017 Dataset}]{
    \includegraphics[width=0.65\columnwidth]{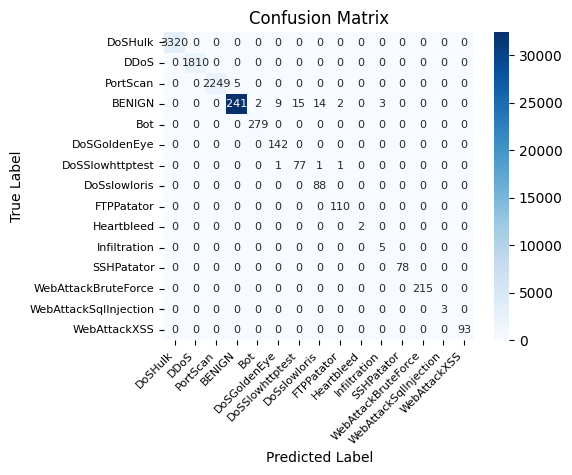}
    \label{fig:cicids2017_confusion}
}
%
\subfigure[{\scriptsize IOTID20 Dataset}]{
    \includegraphics[width=0.65\columnwidth]{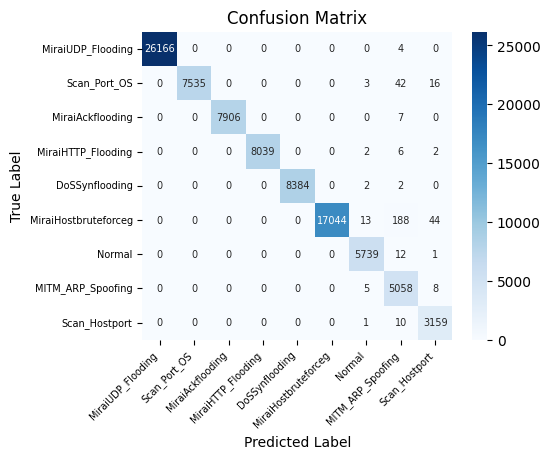}
    \label{fig:iotid20_confusion}
}
\caption{Confusion matrices for the HybridGuard model on different datasets.}
\label{fig:confusion_metrix}
\end{figure*}

\subsection{Performance Based on HybridGuard}

To address the limitations of DualNetShield, particularly in handling data imbalance and detecting minor attacks, we introduced the proposed HybridGuard, which incorporates synthetic data generated using WCGAN-GP.
This synthetic data was combined with the original dataset to form a more balanced and diverse training set. By retraining the model on this enriched dataset, HybridGuard improved the overall performance of the system. 
HybridGuard retains the architecture and model combinations established in DualNetShield, ensuring consistency in the detection process. However, the inclusion of synthetic data greatly enhanced the system's ability to handle data imbalance, allowing it to generalize better and perform more robustly. This improvement is evident across multiple datasets as shown in Table~\ref{tab:table_hybridguard_model}, demonstrating the efficacy of the hybrid approach.

For the UNSW-NB15 dataset, HybridGuard achieved substantial performance gains. The accuracy improved from 94.38\% to 97.07\%, reflecting the enhanced ability to correctly classify network traffic. The F1 Score increased from 95.46\% to 95.87\%, indicating a better balance between precision and recall. Importantly, the false alarm rate (FAR) decreased significantly from 9.22\% to 2.0\%, showcasing the model's improved ability to distinguish between normal and attack traffic. The recall, a critical metric for identifying all attack instances, remained robust at 94.90\%, underscoring the system’s enhanced detection capabilities for less frequent attacks.
Similarly, HybridGuard demonstrated remarkable improvements on the CIC-IDS2017 dataset, where it addressed the high rate of false positives that had previously hindered DualNetShield. The accuracy increased dramatically from 78.89\% to 99.87\%, while the F1 Score rose from 86.36\% to 99.87\%. The precision improved from 98.01\% to 99.89\%, indicating a reduced rate of misclassifications. Additionally, the FAR saw a drastic reduction from 26.09\% to just 0.14\%, further affirming HybridGuard’s effectiveness in distinguishing legitimate traffic from attack traffic. 
For the IoTID20 dataset \cite{ullah2020scheme}, which focuses on IoT-specific attack patterns, HybridGuard achieved notable improvements in false alarm reduction. While metrics such as accuracy (99.6\%), F1 Score (99.5\%), and recall (99.6\%) remained consistent, the FAR significantly decreased from 0.6\% to 0.2\%. This demonstrates HybridGuard’s enhanced ability to minimize false positives, ensuring more reliable detection of IoT-specific minor and rare attacks.

In addition to these improvements, the confusion matrices in Figure \ref{fig:confusion_metrix}, for the three datasets, reveal the effectiveness of HybridGuard in correctly classifying both major and minor attacks. The confusion matrices show a substantial reduction in false positives and false negatives, further underscoring the model's enhanced detection accuracy and reliability.


\begin{figure*}[!t]
\centering
\begin{minipage}{0.48\textwidth}
    \centering
    \subfigure[{\scriptsize HybridGuard}]{
        \includegraphics[height=4.8cm, width=\textwidth]{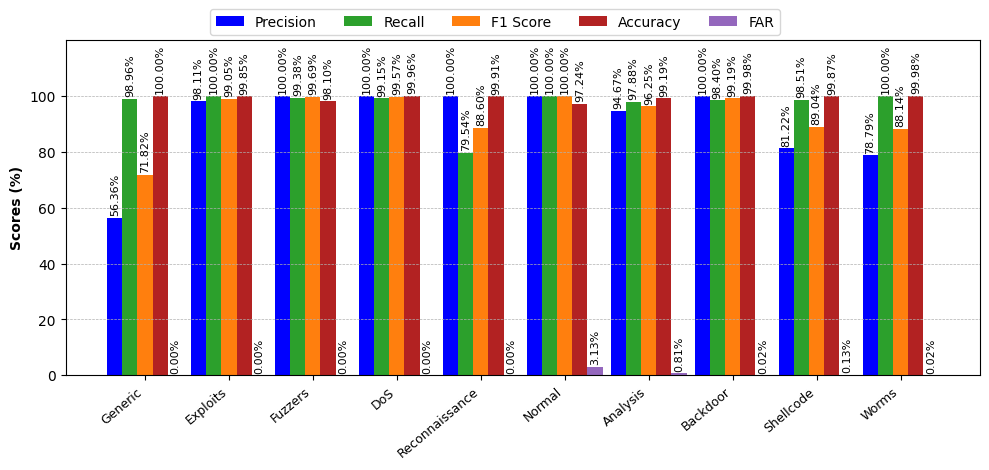}
        \label{fig:hybridguard_MULTIATTACK_unsw-nb15}
    }
\end{minipage}
\hspace{0.02\textwidth} 
\begin{minipage}{0.48\textwidth}
    \centering
    \subfigure[{\scriptsize DualNetShield}]{
        \includegraphics[height=4.8cm, width=\textwidth]{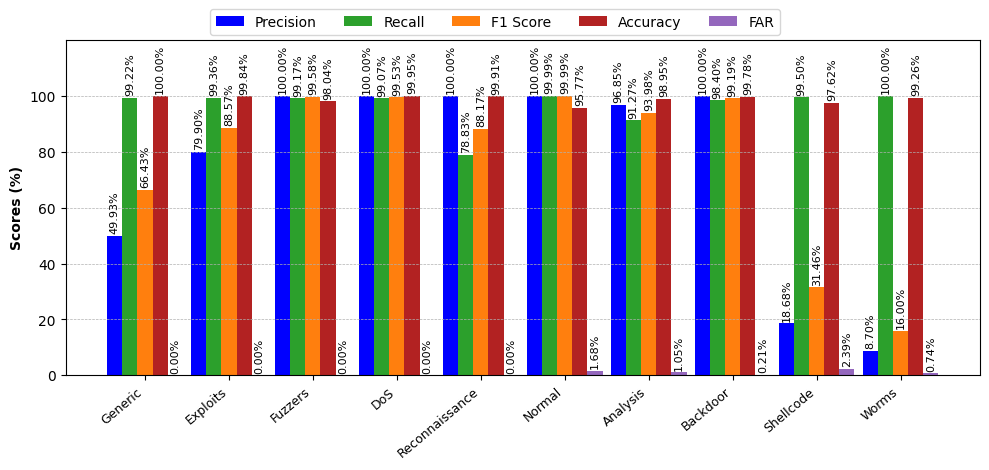}
        \label{fig:dualnetsheild_MULTIATTACK_UNSW-NB15}
    }
\end{minipage}%
\caption{Multiattack classification of UNSW-NB15 dataset: HybridGuard vs. DualNetShield}
\label{fig:multi_attack_comparison_unsw-nb15}
\end{figure*}


\begin{figure*}[!t]
\centering
\begin{minipage}{0.48\textwidth}
    \centering
    \subfigure[{\scriptsize HybridGuard}]{
        \includegraphics[height=4.8cm, width=\textwidth]{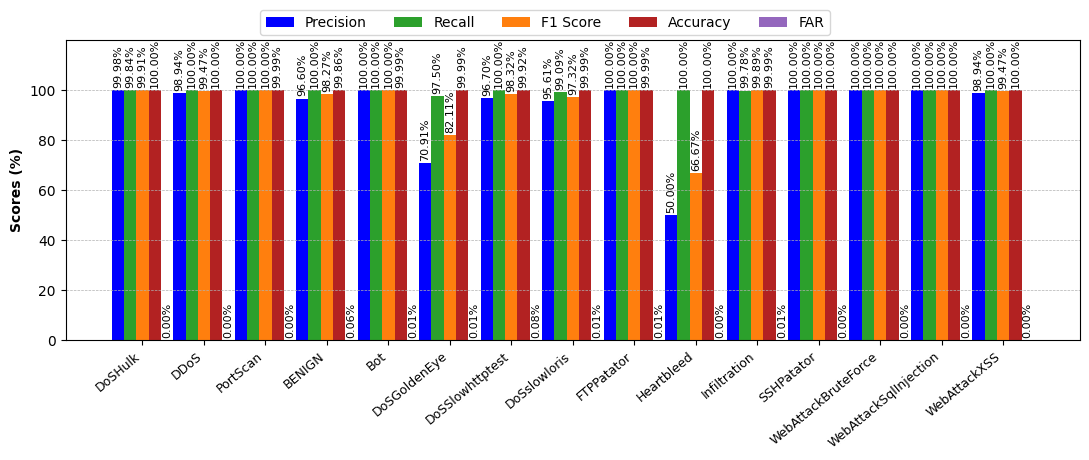}
        \label{fig:hybridguard_MULTIATTACK__cic-ids2017}
    }
\end{minipage}
\hspace{0.02\textwidth} 
\begin{minipage}{0.48\textwidth}
    \centering
    \subfigure[{\scriptsize DualNetShield}]{
        \includegraphics[height=4.8cm, width=\textwidth]{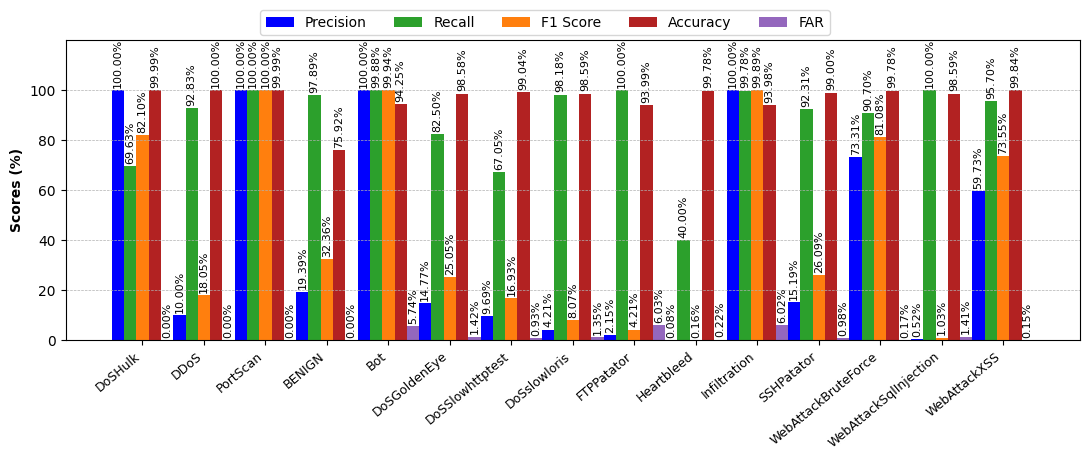}
        \label{fig:dualnetsheild_MULTIATTACK_CIC-IDS2017}
    }
\end{minipage}%
\caption{Multiattack classification of CIC-IDS2017 dataset: HybridGuard vs. DualNetShield.}
\label{fig:multi_attack_comparison_cic-ids2017}
\end{figure*}


\begin{figure*}[!t]
\centering
\begin{minipage}{0.48\textwidth}
    \centering
    \subfigure[{\scriptsize HybridGuard}]{
        \includegraphics[height=4.8cm, width=\textwidth]{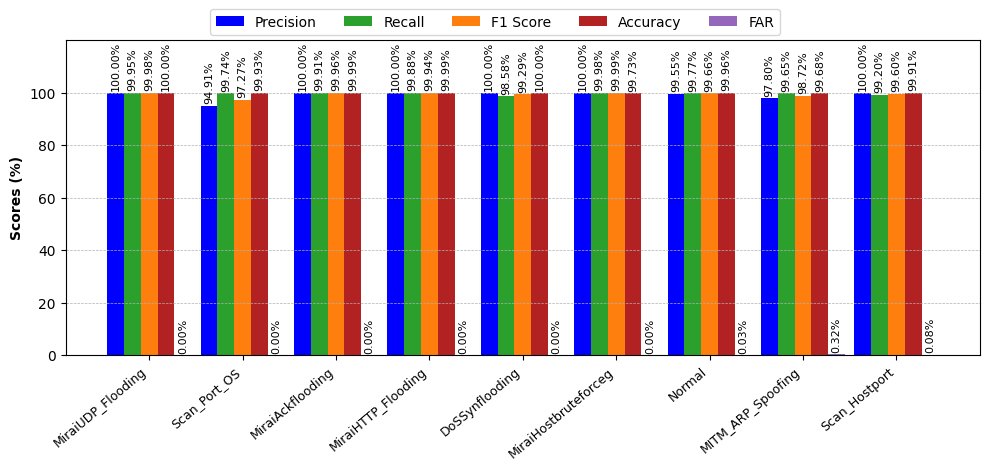}
        \label{fig:hybridguard_MULTIATTACK__IOTID20}
    }
\end{minipage}
\hspace{0.02\textwidth} 
\begin{minipage}{0.48\textwidth}
    \centering
    \subfigure[{\scriptsize DualNetShield}]{
        \includegraphics[height=4.8cm, width=\textwidth]{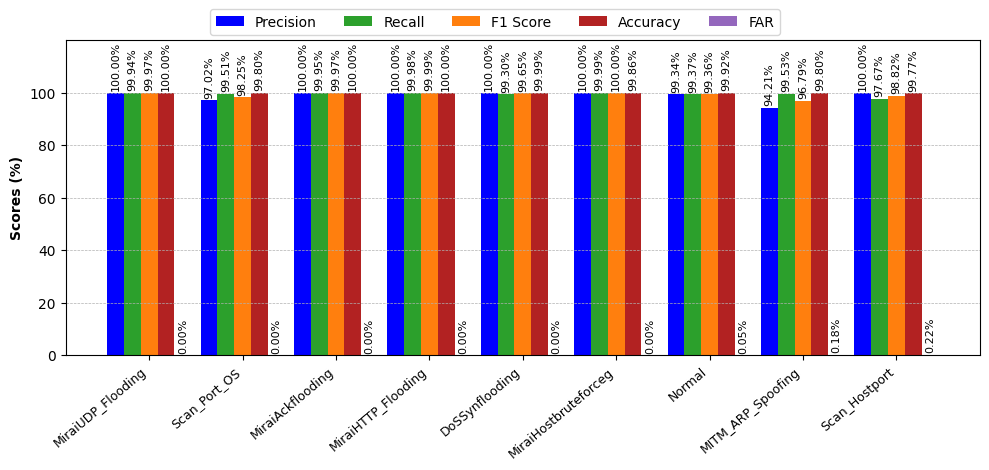}
        \label{fig:dualnetsheild_MULTIATTACK_IOTID20}
    }
\end{minipage}%
\caption{Multiattack classification of IOTID20 dataset: HybridGuard vs. DualNetShield.}
\label{fig:multi_attack_comparison_IOTID20}
\end{figure*}

\subsection{Multiattack Classification Performance: HybridGuard vs. DualNetShield}

HybridGuard successfully addressed the limitations of DualNetShield and demonstrated
superior performance in multi-classification tasks by effectively detecting and 
differentiating between various attack types, both major and minor. 

In Figure~\ref{fig:multi_attack_comparison_unsw-nb15}, the results for the UNSW-NB15 dataset illustrate the superiority of HybridGuard over DualNetShield in both major and minor attack detection. Specifically, Figure~\ref{fig:hybridguard_MULTIATTACK_unsw-nb15} highlights how HybridGuard consistently outperformed in recognizing large-scale attacks such as \textit{Exploits}, \textit{Fuzzers}, and \textit{DoS}. Beyond these dominant categories, HybridGuard demonstrated remarkable sensitivity in detecting less frequent yet critical threats, such as \textit{Shellcode} and \textit{Worms}, which are often overlooked by traditional models. By contrast, Figure~\ref{fig:dualnetsheild_MULTIATTACK_UNSW-NB15} shows the performance of DualNetShield, which, while capable of handling major attacks, displayed weaker detection in rare attack categories, reflecting HybridGuard’s advantage in achieving comprehensive coverage across diverse threat types.  

In Figure~\ref{fig:multi_attack_comparison_cic-ids2017}, which focuses on the CIC-IDS2017 dataset, the robustness of HybridGuard is further reinforced. As shown in Figure~\ref{fig:hybridguard_MULTIATTACK__cic-ids2017}, HybridGuard delivered high precision in identifying dominant threats such as \textit{DDoS} and \textit{PortScan}, both of which are common yet disruptive attack vectors. Moreover, it extended its effectiveness to the accurate recognition of relatively less frequent but sophisticated attacks, including \textit{Web Attacks} and the rare \textit{Heartbleed}. This dual capability translated into outstanding overall accuracy and, importantly, a significantly reduced false alarm rate (FAR), which is a critical metric in real-world deployment scenarios. In comparison, Figure~\ref{fig:dualnetsheild_MULTIATTACK_CIC-IDS2017} depicts the performance of DualNetShield, which, though competent in major categories, exhibited reduced stability and higher FAR, emphasizing HybridGuard’s efficiency in striking a balance between detection depth and operational reliability.  

Finally, in Figure~\ref{fig:multi_attack_comparison_IOTID20}, the evaluation on the IoTID20 dataset demonstrates HybridGuard’s adaptability in IoT-specific intrusion detection tasks. Figure~\ref{fig:hybridguard_MULTIATTACK__IOTID20} reveals that HybridGuard effectively detected large-scale IoT-based attacks such as \textit{MiraiUDP\_Flooding} and \textit{MiraiAckflooding}, which represent critical threats in interconnected device environments. Equally important, HybridGuard showcased strong detection for minor but stealthier intrusions like \textit{Scan\_Hostport} and \textit{MITM\_ARP\_Spoofing}, which pose subtle yet severe risks to IoT ecosystems. This balanced performance underscores HybridGuard’s capability to handle both volumetric and stealthy IoT attacks with equal rigor. Conversely, Figure~\ref{fig:dualnetsheild_MULTIATTACK_IOTID20} highlights the limitations of DualNetShield, which was comparatively weaker in minor attack detection, thus validating HybridGuard’s enhanced adaptability and comprehensive protection across IoT scenarios.

%
%

These results, in Figures~\ref{fig:multi_attack_comparison_unsw-nb15}, \ref{fig:multi_attack_comparison_cic-ids2017}, and \ref{fig:multi_attack_comparison_IOTID20} illustrate the superior performance of HybridGuard in addressing data imbalance, enhancing detection rates, and reducing false positives. By combining WCGAN-GP with DualNetShield, HybridGuard offers a robust solution for multi-class attack classification tasks capable of handling both frequent and rare attack types effectively.


\section{Conclusion and Future Work} 
\label{sec:conclusion}
This paper introduces the HybridGuard framework, a new AI-based system for detecting network intrusions specifically designed to solve the problem of imbalanced data. Using the WCGAN-GP model to create synthetic data for underrepresented attack types, we improved the training process and made detection more accurate, especially for minority attack classes. The DualNetShield model, part of the HybridGuard framework, showed strong performance in identifying both major and minor attacks.

The experiments, UNSW-NB15, CIC-IDS2017, and IoTID20 datasets confirmed that HybridGuard outperformed existing machine learning and deep learning models. For UNSW-NB15 and CIC-IDS2017, HybridGuard achieved accuracy rates of 97.09\% and 99.87\%, respectively, while significantly reducing the False Alarm Rate (FAR). On the IoTID20 dataset, which focuses on IoT-specific attack patterns, HybridGuard maintained a high accuracy of 99.6\%, effectively detecting rare and minority attack types. Moreover, the FAR on the IoTID20 dataset dropped from 0.6\% to 0.2\%, highlighting HybridGuard’s exceptional ability to minimize false positives in IoT-specific environments. These results demonstrate HybridGuard's effectiveness in accurately detecting a wide range of network attacks, making it suitable for real-world cybersecurity applications.

In the future, we aim to enhance HybridGuard for broader distributed network challenges. This includes integrating it into federated learning for decentralized adaptability, enhancing threat detection with advanced AI models, and evaluating its scalability on large-scale IoT datasets in edge computing. Additionally, we will study adversarial attacks on AI-based intrusion detection and develop defenses to strengthen HybridGuard against such threats, making it a more robust solution for modern networked systems.

\section*{Acknowledgment} 
This work was supported by the National Science and Technology Council (NSTC), Taiwan, under Grants 112-2221-E-011-057- and 114-2221-E-011-137-.

\subsubsection*{CRediT authorship contribution statement}

Binayak Kar: Problem definition, Supervision, Validation, analysis, Writing–review \& editing. 
Ujjwal Sahu: Problem definition, Validation, analysis, Algorithm design, Software, Experimental study, Writing. 
Ciza Thomas: Supervision, Validation, Review. 
Jyoti Prakash Sahoo: Algorithm design, Paper revision.

\subsubsection*{Declaration of interests} 

The authors declare that they have no known competing financial interests or personal relationships that could have appeared to influence the work reported in this paper.

\bibliographystyle{unsrtnat}
\bibliography{HybridGuard.bib}

\end{document}